\documentclass[12pt, draftclsnofoot, onecolumn]{IEEEtran}
\usepackage{amsthm}
\usepackage{multirow}
\usepackage{color}

\usepackage[T1]{fontenc}
\usepackage{algorithm}
\usepackage{algorithmic}
\ifCLASSINFOpdf
\else
\fi
\usepackage{amsmath}
\usepackage{amsmath,bm}
\usepackage{graphicx}
\usepackage{amsmath,amssymb,amsthm,mathrsfs,amsfonts,dsfont}
\usepackage{subfigure}
\usepackage{array}
\begin{document}
%
\title{Two-Dimensional AoD and AoA Acquisition for Wideband mmWave Systems with Cross-Polarized MIMO}
%
%
%

\author{Dalin Zhu,
        Junil Choi,
        and~Robert~W.~Heath~Jr.
\thanks{Dalin Zhu and Robert W. Heath Jr. are with the Department
of Electrical and Computer Engineering, The University of Texas at Austin, Austin,
TX, 78712 USA, e-mail: \{dalin.zhu, rheath\}@utexas.edu.

Junil Choi is with the Department of Electrical Engineering, Pohang University of Science and Technology (POSTECH), Pohang, Gyeongbuk 37673 Korea, e-mail: junil@postech.ac.kr.

Parts of this work have been accepted for presentation at IEEE Globecom 2016 \cite{dzglobe}. This research was partially supported by the U.S. Department of Transportation through the Data-Supported Transportation Operations and Planning (D-STOP) Tier 1 University Transportation Center, and by a gift from Huawei Technologies.}}

\maketitle

\begin{abstract}
In this paper, a novel two-dimensional super-resolution angle-of-departure (AoD) and angle-of-arrival (AoA) estimation technique is proposed for wideband millimeter-wave multiple-input multiple-output systems with cross-polarized antenna elements. The key ingredient of the proposed method is to form custom designed beam pairs, and devise an invertible function of the AoD/AoA to be estimated from the corresponding beam pairs. Further, a new multi-layer reference signal structure is developed for the proposed method to facilitate angle estimation for wideband channels with cross-polarized antenna elements. To facilitate feedback in closed-loop frequency division duplexing systems, a novel differential feedback strategy is proposed aiming at feedback reduction for the two-dimensional angle estimation. Numerical results demonstrate that by using the proposed method, good azimuth/elevation AoD and AoA estimation performance can be achieved under different levels of signal-to-noise ratio, channel conditions, and antenna array configurations.
\end{abstract}


%
\IEEEpeerreviewmaketitle

\allowdisplaybreaks

\section{Introduction}

Due to the large available spectral channels, ultra-high data rates can be supported by the millimeter-wave (mmWave) band for wireless local area networks \cite{ieeewlan} and fifth generation (5G) cellular networks \cite{jerrypikhan}-\nocite{rhsp}\cite{pchj}. The small carrier wavelengths at mmWave frequencies enable synthesis of compact antenna arrays, providing large beamforming gains to support sufficient link margin \cite{mmwavebook}. Cross-polarized antenna systems, discussed in \cite{dual0,dual1}, can also be incorporated into the mmWave systems allowing a large number of antennas to be deployed with a small form factor. In addition to this space efficiency, high rank data streams can be multiplexed across the polarizations by taking advantage of cross-pole decoupling in the channel. In a cross-polarized mmWave multiple-input multiple-output (MIMO) system, it is useful to acquire accurate angle-of-departure (AoD) and angle-of-arrival (AoA) of the channels, with which highly directional beams can be formed for data communications. To develop practical angle estimation algorithms, the fact that the AoD and AoA are the same for both polarization domains can be leveraged, while at the same time recognizing that wideband channels should also be frequency selective.

MmWave specific narrowband channel estimation has been studied in \cite{ahmedce}-\nocite{uva1}\nocite{rmr}\cite{singh2}. The main challenge in mmWave systems is that the use of analog and hybrid beamforming spatially compresses the high dimensional channel as experienced in the digital domain. In \cite{ahmedce,uva1}, algorithms that exploit channel sparsity were proposed for channel estimation based on compressed sensing. An open-loop channel estimation strategy was developed in \cite{rmr}, in which the estimation algorithm was independent of the hardware constraints. In \cite{singh2}, the grid-of-beams (GoB) based approach, also exploited in UMTS Release 6 \cite{3gr6}, was employed to acquire the channel directional information via beam training for mmWave systems. In the GoB methods, the best combinations of analog transmit and receive beams, which characterize the channel's AoD and AoA, were obtained. A large amount of training is needed for the techniques proposed in \cite{ahmedce}-\nocite{uva1}\nocite{rmr}\cite{singh2}, which may be computationally prohibitive and also requires large overhead. Further, only quantized angle estimation can be achieved via the compressed sensing \cite{ahmedce,rmr} and grid-of-beams \cite{singh2} based methods. With a small codebook or dictionary, the angle estimation resolution is limited, which in turn, provides an error floor in the overall estimation performance. Also, the aforementioned techniques applied only for narrowband mmWave channels.

Recently, channel estimation for wideband mmWave channels was considered in \cite{wb1,wb2}. In \cite{wb1}, orthogonal frequency-division multiplexing (OFDM) was assumed, while the technique developed in \cite{wb2} was suitable for both single-carrier and multi-carrier modulations. The hybrid architecture was exploited in both \cite{wb1} and \cite{wb2}, but neither approach considered polarization. In \cite{jssldl}, the cross-polarization was considered for the first time when performing the beam training in narrowband mmWave systems. The channel estimation techniques proposed in \cite{wb1}-\nocite{wb2}\cite{jssldl} focused on either the wideband channels or cross-polarization, but not the both.

To better trade-off the estimation overhead and resolution, we propose to estimate the channel's azimuth/elevation AoD and AoA through an auxiliary beam pair (ABP) design for cross-polarized MIMO in wideband mmWave systems. Construction of pairs of beams was previously employed in monopulse radar systems to improve the AoA estimation accuracy \cite{radar0}. In amplitude monopulse radar, a sum beam and a difference beam form a pair such that the difference beam steers a null towards the boresight angle of the sum beam. By comparing the relative amplitude of the pulse in the pair of two beams, the direction of the target can be determined with accuracy dependent on the received signal-to-noise ratio (SNR). Directly applying the technique for monopulse radar to communications systems will yield large overhead as the difference beams are only used for assisting the sum beams to conduct the angle estimation, not providing any angular coverage. This is because the angular coverage provided by the monopulse beam pair is approximately the same as the half-power beamwidth of the corresponding sum beam. In our previous work in \cite{dztrans}, the idea of auxiliary beam pair design was exploited to acquire high-resolution one-dimensional angle information for mmWave. The proposed algorithms, however, are only suited for uniform linear arrays (ULAs) with direct quantization and feedback strategies, which may result in excessive feedback overhead if two-dimensional angle estimation is conducted. In our preliminary work in \cite{dzglobe}, we extended our work in \cite{dztrans} to two-dimensional arrays, but only considered a single-path narrowband channel. Our prior work in \cite{dzglobe,dztrans} mainly focused on the feasibility of the proposed auxiliary beam pairs design in estimating the channel directional information, and does not apply to wideband channels or channels with polarization.

In this paper, we propose an auxiliary beam pair based channel estimators, which works for wideband channels and polarization. To facilitate practical realization of the proposed design, we provide several feedback options and guidelines on pilot designs. The main contributions of the paper are summarized as follows:
\begin{itemize}
  \item \emph{Auxiliary beam pair enabled joint azimuth/elevation AoD and AoA estimation}. In the proposed design approach, for a given azimuth (elevation) steering direction, two beams are formed as an elevation (azimuth) transmit auxiliary beam pair such that they steer towards successive angular directions in the elevation (azimuth) domain. The formed elevation and azimuth transmit auxiliary beam pairs are used to estimate the elevation and azimuth transmit spatial frequencies, and therefore, the corresponding AoDs. The estimation of AoA is similarly performed by forming receive auxiliary beam pairs.
  \item \emph{Angle estimation with cross-polarization}. With cross-polarized antenna elements, the analog beams within each auxiliary beam pair are formed across both vertical and horizontal polarization domains. Various auxiliary beam pair and polarization mapping strategies are discussed, and their impacts on the system design are also evaluated.
  \item \emph{Quantization options}. For frequency division duplexing (FDD), a new differential feedback option is developed for the proposed auxiliary beam pair design. In contrast to the direct quantization we developed in \cite{dztrans}, the proposed differential quantization method is able to reduce the feedback overhead for two-dimensional angle acquisition.
  \item \emph{Multi-layer pilot structure}. Building on the single-path solution, to facilitate the multi-path angle acquisition, multiple RF chains are employed such that multiple auxiliary beam pairs are probed simultaneously. Accordingly, a new multi-layer pilot structure is proposed. In the first layer, the auxiliary beam pair specific pilot design is developed to distinguish between simultaneously probed auxiliary beam pairs. In the second layer, the paired-beam specific pilot is proposed to independently decouple the beams in the same auxiliary beam pair.
\end{itemize}
Our proposed angle estimation algorithm is evaluated by simulations under realistic assumptions. The numerical results reveal that the proposed design approach is capable of providing high-resolution azimuth/elevation AoD and AoA estimation under various SNR levels and channel conditions subject to moderate amounts of feedback and training overheads.

The rest of this paper is organized as follows. Section II specifies the system and channel models. Section III describes the principles and procedures of the proposed auxiliary beam pair design in estimating the two-dimensional angles using the multi-layer pilot structure. Section IV illustrates several practical issues of implementing the proposed algorithm, including duplexing, feedback options, and codebook design. Section V shows numerical results to validate the effectiveness of the proposed technique. Finally, Section VI draws our conclusions.

\textbf{Notations}: $\bm{A}$ is a matrix; $\bm{a}$ is a vector; $a$ is a scalar; $|a|$ is the magnitude of the complex number
$a$. $(\cdot)^{\mathrm{T}}$ and $(\cdot)^{*}$ denote transpose and conjugate transpose; $\|\bm{A}\|_{\mathrm{F}}$ is the Frobenius norm of $\bm{A}$ and $\det(\bm{A})$ is its determinant; $\left[\bm{A}\right]_{:,j}$ is the $j$-th column of $\bm{A}$; $\left[\bm{A}\right]_{i,j}$ is the $(i,j)$-th entry of $\bm{A}$; $\mathrm{tr}(\bm{A})$ is the trace of $\bm{A}$; $\bm{I}_{N}$ is the $N\times N$ identity matrix; $\bm{1}_{M\times N}$ represents the $M\times N$ matrix whose entries are all ones; $\bm{0}_{N}$ denotes the $N\times 1$ vector whose entries are all zeros; $\mathcal{N}_{c}(\bm{a},\bm{A})$ is a complex Gaussian vector with mean $\bm{a}$ and covariance $\bm{A}$; $\mathbb{E}[\cdot]$ is used to denote expectation, $\odot$ is the Hadamard product, and $\otimes$ is the Kronecker product; $\mathrm{sign}(\cdot)$ extracts the sign of a real number; $\mathrm{gcd}(\cdot)$ stands for the greatest common divisor; $\mathrm{diag}(\cdot)$ is the diagonalization operation.

\section{System and Channel Models}
\begin{figure}
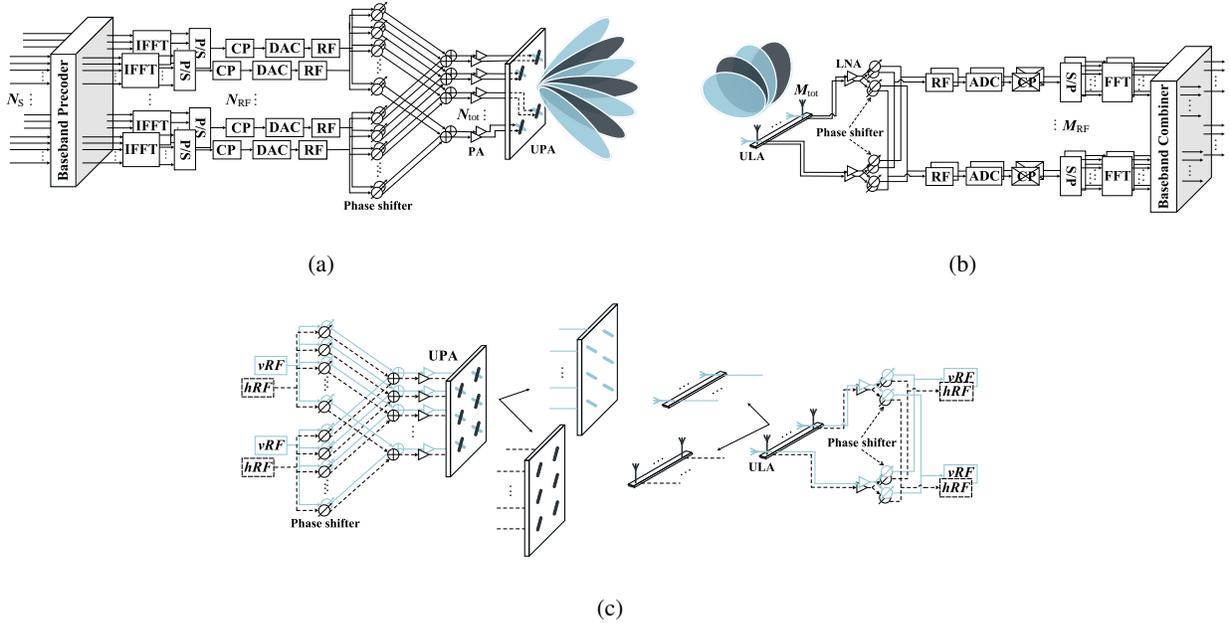

\centering
\subfigure[]{%
\includegraphics[width=3.35in]{transmitter_shared_xpol_rev.pdf}
\label{fig:subfigure1}}
\quad
\subfigure[]{%
\includegraphics[width=2.8in]{receiver_shared_xpol.pdf}
\label{fig:subfigure2}}
\quad
\subfigure[]{%
\includegraphics[width=3.90in]{enlarge_UPA_xpol.pdf}
\label{fig:subfigure2}}
\caption{(a) Partially shared-array architecture is employed at the BS with $N_{\mathrm{RF}}$ RF chains and a total number of $N_{\mathrm{tot}}$ cross-polarized antenna elements. $N_{\mathrm{S}}$ data streams are transmitted. (b) Partially shared-array architecture is employed at the UE with $M_{\mathrm{RF}}$ RF chains and a total number of $M_{\mathrm{tot}}$ cross-polarized antenna elements. (c) Detailed polarization and RF chain mapping for the proposed partially shared-array architecture.}
\label{fig:figure}
\end{figure}
In this section, hybrid precoding MIMO-OFDM system and wideband channel models with both co-polarization and cross-polarization are presented.
\subsection{System model}
We consider a precoded MIMO-OFDM system with $N$ subcarriers and a hybrid precoding transceiver structure as shown in Figs.~1(a) and 1(b), in which a base station (BS) equipped with $N_{\mathrm{tot}}$ transmit antennas and $N_{\mathrm{RF}}$ RF chains transmits $N_{\mathrm{S}}$ data streams to a user equipment (UE) equipped with $M_{\mathrm{tot}}$ receive antennas and $M_{\mathrm{RF}}$ RF chains. Here, $N_{\mathrm{S}}\leq M_{\mathrm{RF}}\leq N_{\mathrm{RF}}$. Cross-polarized antenna elements are equipped at both the BS and UE such that all antenna elements are evenly divided into vertically and horizontally polarized antennas. Both the BS and UE are equipped with partially shared-array antenna architectures. In Fig.~1(c), the exact mapping between the RF chains and the cross-polarized antenna elements for the developed partially shared-array architecture is illustrated. Specifically, all RF chains are first evenly grouped into sets of vertical and horizontal RF chains, denoted as \emph{vRF} and \emph{hRF} in Fig.~1(c). All vertically (horizontally) polarized antenna elements are jointly controlled by all corresponding vertical (horizontal) RF chains sharing the same network of phase shifters. Analog beamforming is independently performed from each polarization domain via the corresponding vertical or horizontal RF chain. In addition, a uniform planar array (UPA) is equipped at the BS. For simplicity, the ULA is assumed at the UE side for better illustration of the design procedures for the two-dimensional AoD estimation via the proposed method, though the algorithm can be equally applied to the two-dimensional AoA estimation if the UE is equipped with the UPA.

Let $\bm{s}[k]$ denote an $N_{\mathrm{S}}\times 1$ baseband transmit symbol vector such that $\mathbb{E}\left[\bm{s}[k]\bm{s}^{*}[k]\right]=\bm{I}_{N_{\mathrm{S}}}$, $k=0,\cdots,N-1$. The data symbol vector $\bm{s}[k]$ is first precoded using an $N_{\mathrm{RF}}\times N_{\mathrm{S}}$ digital baseband precoding matrix $\bm{F}_{\mathrm{BB}}[k]$ on the $k$-th subcarrier, resulting in $\bm{d}[k]=\left[d_1[k],\cdots,d_{N_{\mathrm{RF}}}[k]\right]^{\mathrm{T}}=\bm{F}_{\mathrm{BB}}[k]\bm{s}[k]$. The data symbol blocks are then transformed to the time-domain via $N_{\mathrm{RF}}$, $N$-point IFFTs, generating discrete-time signal as $\bm{x}_{n_{\mathrm{R}}}[k]=\sum_{k'=1}^{N}d_{n_{\mathrm{R}}}[k']e^{\frac{j2\pi k'}{N}k}$, where $n_{\mathrm{R}}=1,\cdots,N_{\mathrm{RF}}$ and $k=0,\cdots,N-1$. Before applying an $N_{\mathrm{tot}}\times N_{\mathrm{RF}}$ wideband analog precoding matrix $\bm{F}_{\mathrm{RF}}$, a cyclic prefix (CP) with length $D$ is added to the data symbol blocks such that $D$ is greater than or equal to the maximum delay spread of the multi-path channel. Denote by $\bm{x}[k_{\mathrm{c}}]=\left[\bm{x}_{1}[k_{\mathrm{c}}],\cdots,\bm{x}_{N_{\mathrm{RF}}}[k_{\mathrm{c}}]\right]^{\mathrm{T}}$, where $k_{\mathrm{c}}=N-D,\cdots,N-1,0,\cdots,N-1$ due to the insertion of the CP. The discrete-time transmit signal model is then expressed as $\bm{x}_{\mathrm{cp}}[k_{\mathrm{c}}] = \bm{F}_{\mathrm{RF}}\bm{x}[k_{\mathrm{c}}]$. As the analog precoder is implemented using analog phase shifters, $\left[\left[\bm{F}_{\mathrm{RF}}\right]_{:,n_{\mathrm{R}}}\left[\bm{F}_{\mathrm{RF}}\right]_{:,n_{\mathrm{R}}}^{*}\right]_{i,i}=\frac{1}{N_{\mathrm{tot}}}$ is satisfied, where $n_{\mathrm{R}}=1,\cdots,N_{\mathrm{RF}}$ and $i=1,\cdots,N_{\mathrm{tot}}$, i.e., all elements of $\bm{F}_{\mathrm{RF}}$ have equal norm. To further ensure the total power constraint, $\sum_{k=1}^{N}\left\|\bm{F}_{\mathrm{RF}}\bm{F}_{\mathrm{BB}}[k]\right\|_{\mathrm{F}}^{2}=N_{\mathrm{S}}$ is required.

At the UE side, after combining with an $M_{\mathrm{tot}}\times M_{\mathrm{RF}}$ analog combining matrix $\bm{W}_{\mathrm{RF}}$, the CP is removed. After transforming the received data symbols from time-domain to frequency-domain via $M_{\mathrm{RF}}$, $N$-point FFTs, an $M_{\mathrm{RF}}\times N_{\mathrm{S}}$ digital baseband combining matrix $\bm{W}_{\mathrm{BB}}[k]$ is applied on subcarrier $k$. The discrete-time received signal can then be expressed as
\begin{equation}\label{resigsub}
\bm{y}[k]=\bm{W}^{*}_{\mathrm{BB}}[k]\bm{W}_{\mathrm{RF}}^{*}\bm{H}[k]\bm{F}_{\mathrm{RF}}\bm{F}_{\mathrm{BB}}[k]\bm{s}[k]+\bm{W}_{\mathrm{BB}}^{*}[k]\bm{W}_{\mathrm{RF}}^{*}\bm{n}[k].
\end{equation}
Here, $\bm{n} \sim \mathcal{N}_{c}(\bm{0}_{M_{\mathrm{tot}}},\sigma^{2}\bm{I}_{M_{\mathrm{tot}}})$ and $\sigma^2=1/\gamma$, where $\gamma$ represents the target SNR.
\subsection{Channel models}
To characterize the angular sparsity and frequency selectivity of the wideband mmWave channels, a spatial geometric channel model is employed. Spatial geometric channel models have been adopted in Long-Term Evolution (LTE) systems for various deployment scenarios \cite{ltech}. In this paper, a simplified version of those developed in \cite{ltech} is used by assuming that the channel has $N_{\mathrm{r}}$ paths, and each path $r$ has azimuth and elevation AoDs $\phi_{r}$, $\theta_{r}$, and AoA $\psi_{r}$. Note that the numerical results in Section V are obtained by using more realistic channel model assumptions, which are similar to those proposed in \cite{ltech}, though the simplified model is employed for better illustration of the design principle for the proposed algorithm.

\subsubsection{Co-polarized channel model}
Let $p(\tau)$ denote the pulse shaping function for $T_{\mathrm{s}}$-spaced signaling at $\tau$ seconds. Assuming only co-polarized antenna elements at both the BS and UE, the time-domain delay-$d$ MIMO channel matrix can be expressed as
\begin{equation}\label{delayd}
\bm{G}[d]=\sum_{r=1}^{N_{\mathrm{r}}}g_{r}p\left(dT_{\mathrm{s}}-\tau_{r}\right)\bm{a}_{\mathrm{r}}(\psi_{r})\bm{a}_{\mathrm{t}}^{*}(\theta_{r},\phi_{r}),
\end{equation}
where $g_{r}$ represents the complex path gain of path-$r$, $\bm{a}_{\mathrm{r}}(\cdot)\in\mathbb{C}^{M_{\mathrm{tot}}\times1}$ and $\bm{a}_{\mathrm{t}}(\cdot)\in\mathbb{C}^{N_{\mathrm{tot}}\times1}$ correspond to the receive and transmit array response vectors. Specifically, assuming that the UPA is employed by the BS in the $\mathrm{xy}$-plane with $N_{\mathrm{x}}$ and $N_{\mathrm{y}}$ elements on the $\mathrm{x}$ and $\mathrm{y}$ axes,
\begin{eqnarray}\label{upaco}
\bm{a}_{\mathrm{t}}(\theta_{r},\phi_{r})&=&\frac{1}{\sqrt{N_{\mathrm{tot}}}}\Big[1, e^{j\frac{2\pi}{\lambda}d_{\mathrm{tx}}\sin(\theta_{r})\cos(\phi_{r})},\cdots,e^{j\frac{2\pi}{\lambda}\left(N_{\mathrm{x}}-1\right)d_{\mathrm{tx}}\sin(\theta_{r})\cos(\phi_{r})}, e^{j\frac{2\pi}{\lambda}d_{\mathrm{tx}}\sin(\theta_{r})\sin(\phi_{r})},\nonumber\\
&&\cdots,e^{j\frac{2\pi}{\lambda}\left(\left(N_{\mathrm{x}}-1\right)d_{\mathrm{tx}}\sin(\theta_{r})\cos(\phi_{r})+\left(N_{\mathrm{y}}-1\right)d_{\mathrm{ty}}\sin(\theta_{r})\sin(\phi_{r})\right)}\Big]^{\mathrm{T}},
\end{eqnarray}
where $N_{\mathrm{tot}}=N_{\mathrm{x}}N_{\mathrm{y}}$, $\lambda$ represents the wavelength corresponding to the operating carrier frequency, $d_{\mathrm{tx}}$ and $d_{\mathrm{ty}}$ are the inter-element distances of the co-polarized antenna elements on the $\mathrm{x}$ and $\mathrm{y}$ axes. Further, since the ULA is employed by the UE,
\begin{equation}\label{ulaaaco}
\bm{a}_{\mathrm{r}}(\psi_{r})=\frac{1}{\sqrt{M_{\mathrm{tot}}}}\big[1, e^{j\frac{2\pi}{\lambda}d_{\mathrm{r}}\sin(\psi_{r})},\cdots,e^{j\frac{2\pi}{\lambda}d_{\mathrm{r}}\left(M_{\mathrm{tot}}-1\right)\sin(\psi_{r})} \big]^{\mathrm{T}},
\end{equation}
where $d_{\mathrm{r}}$ denotes the inter-element distance between the co-polarized receive antenna elements. The channel frequency response matrix on subcarrier $k$ is
\begin{equation}\label{wbfrnp}
\bm{H}[k]=\sum_{r=1}^{N_{\mathrm{r}}}g_{r}\rho_{\tau_{r}}[k]\bm{a}_{\mathrm{r}}(\psi_{r})\bm{a}_{\mathrm{t}}^{*}(\theta_{r},\phi_{r}),
\end{equation}
where $\rho_{\tau_{r}}[k]=\sum_{d=0}^{D-1}p\left(dT_{\mathrm{s}}-\tau_{r}\right)e^{-\frac{j2\pi kd}{N}}$.

\subsubsection{Cross-polarized channel model}
The basic model in (\ref{wbfrnp}), needs further modification to account for the use of cross-polarized antennas. A cross polar discrimination (XPD) value is incorporated, which implies the ability to distinguish between different polarized antennas \cite{jssldl}. The difference in orientation of the polarized antenna groups is further characterized by a Givens rotation matrix. To be more specific, denote by $g_{r}^{\mathrm{ab}}$ the complex path gain from polarization $\mathrm{a}$ to $\mathrm{b}$ for path-$r$, the channel model in (\ref{wbfrnp}) can be rewritten as \cite{gcdc}-\nocite{jssldl}\cite{scll}
\begin{eqnarray}\label{wbfrp}
\bm{H}[k]=\sum_{r=1}^{N_{\mathrm{r}}}\rho_{\tau_{r}}[k]\left(\bm{X}_{\chi}\odot\left(\left[\begin{array}{cc}
                                                                                      g_{r}^{\mathrm{vv}} & g_{r}^{\mathrm{vh}}  \\
                                                                                      g_{r}^{\mathrm{hv}}  & g_{r}^{\mathrm{hh}}
                                                                                    \end{array}
\right]\otimes\bigg(\bm{a}_{\mathrm{r}}(\psi_{r})\bm{a}_{\mathrm{t}}^{*}(\theta_{r},\phi_{r})\bigg)\right)\right)\bm{R}_{\varsigma},
\end{eqnarray}
where
\begin{equation}
\bm{X}_{\chi}=\sqrt{\frac{1}{1+\chi}}\left[\begin{array}{cc}
                                      1 & \sqrt{\chi} \\
                                      \sqrt{\chi} & 1
                                    \end{array}
\right]\otimes\bm{1}_{\frac{M_{\mathrm{tot}}}{2}\times \frac{N_{\mathrm{tot}}}{2}},
\end{equation}
represents the power imbalance between the polarizations with the parameter $\chi$, defining the reciprocal of the XPD, and,
\begin{equation}
\bm{R}_{\varsigma}=\left[\begin{array}{cc}
                       \cos(\varsigma) & -\sin(\varsigma) \\
                       \sin(\varsigma) & \cos(\varsigma)
                     \end{array}
\right]\otimes\bm{I}_{\frac{N_{\mathrm{tot}}}{2}},
\end{equation}
denotes the Givens rotation matrix for a mismatch angle of $\varsigma$. Assuming $N_{\mathrm{tot}}=N_{\mathrm{x}}N_{\mathrm{y}}$, $\bm{a}_{\mathrm{t}}(\cdot)\in\mathbb{C}^{\frac{N_{\mathrm{tot}}}{2}\times1}$ and $\bm{a}_{\mathrm{r}}(\cdot)\in\mathbb{C}^{\frac{M_{\mathrm{tot}}}{2}\times1}$ are expressed as
\begin{eqnarray}\label{upa}
\bm{a}_{\mathrm{t}}(\theta_{r},\phi_{r})&=&\frac{1}{\sqrt{N_{\mathrm{tot}}/2}}\Big[1, e^{j\frac{2\pi}{\lambda}d_{\mathrm{tx}}\sin(\theta_{r})\cos(\phi_{r})},\cdots,e^{j\frac{2\pi}{\lambda}\left(\frac{N_{\mathrm{x}}}{2}-1\right)d_{\mathrm{tx}}\sin(\theta_{r})\cos(\phi_{r})}, e^{j\frac{2\pi}{\lambda}d_{\mathrm{tx}}\sin(\theta_{r})\sin(\phi_{r})},\nonumber\\
&&\cdots,e^{j\frac{2\pi}{\lambda}\left(\left(\frac{N_{\mathrm{x}}}{2}-1\right)d_{\mathrm{tx}}\sin(\theta_{r})\cos(\phi_{r})+\left(\frac{N_{\mathrm{y}}}{2}-1\right)d_{\mathrm{ty}}\sin(\theta_{r})\sin(\phi_{r})\right)}\Big]^{\mathrm{T}},
\end{eqnarray}
\begin{equation}\label{ulaaa}
\bm{a}_{\mathrm{r}}(\psi_{r})=\frac{1}{\sqrt{M_{\mathrm{tot}}/2}}\big[1, e^{j\frac{2\pi}{\lambda}d_{\mathrm{r}}\sin(\psi_{r})},\cdots,e^{j\frac{2\pi}{\lambda}d_{\mathrm{r}}\left(\frac{M_{\mathrm{tot}}}{2}-1\right)\sin(\psi_{r})} \big]^{\mathrm{T}}.
\end{equation}
As can be seen from (\ref{upaco}), (\ref{ulaaaco}), (\ref{upa}) and (\ref{ulaaa}), the dimensionality of $\bm{a}_{\mathrm{t}}(\cdot)$ or $\bm{a}_{\mathrm{r}}(\cdot)$ for the cross-polarized antenna array is one half of that for the co-polarized antenna array.
\section{Auxiliary Beam Pair Enabled Two-Dimensional Angle Estimation}
In this section, our proposed design approach is first described assuming a simple narrowband single-path channel using a single RF chain without considering cross-polarization. Afterwards, a means of implementing the proposed method in estimating wideband multi-path angle components using multiple RF chains assuming cross-polarized antenna elements is presented.

\subsection{Narrowband single-path angle estimation using single RF chain without cross-polarization}
For narrowband single-path channels with co-polarized antenna elements, the channel model in (\ref{wbfrnp}) can be simplified as $\bm{H}=g\bm{a}_{\mathrm{r}}(\psi)\bm{a}_{\mathrm{t}}^{*}(\theta,\phi)$, where the path and subcarrier indices are dropped. As only co-polarized antenna setup is assumed in this subsection, $\bm{a}_{\mathrm{r}}(\psi)\in\mathbb{C}^{M_{\mathrm{tot}}\times 1}$ and $\bm{a}_{\mathrm{t}}(\theta,\phi)\in\mathbb{C}^{N_{\mathrm{tot}}\times 1}$. Denote by $\mu_{\mathrm{x}}=\frac{2\pi}{\lambda}d_{\mathrm{tx}}\sin(\theta)\cos(\phi)$ and $\mu_{\mathrm{y}}=\frac{2\pi}{\lambda}d_{\mathrm{ty}}\sin(\theta)\sin(\phi)$, which can be interpreted as the elevation and azimuth transmit spatial frequencies. We further define two vectors $\bm{a}_{\mathrm{tx}}(\mu_{\mathrm{x}})\in\mathbb{C}^{N_{\mathrm{x}}\times 1}$ and $\bm{a}_{\mathrm{ty}}(\mu_{\mathrm{y}})\in\mathbb{C}^{N_{\mathrm{y}}\times 1}$ as
\begin{eqnarray}
\bm{a}_{\mathrm{tx}}(\mu_{\mathrm{x}})=\frac{1}{\sqrt{N_{\mathrm{x}}}}\left[1, e^{j\mu_{\mathrm{x}}},\cdots, e^{j\left(N_{\mathrm{x}}-1\right)\mu_{\mathrm{x}}} \right]^{\mathrm{T}},\bm{a}_{\mathrm{ty}}(\mu_{\mathrm{y}})=\frac{1}{\sqrt{N_{\mathrm{y}}}}\left[1, e^{j\mu_{\mathrm{y}}},\cdots, e^{j\left(N_{\mathrm{y}}-1\right)\mu_{\mathrm{y}}} \right]^{\mathrm{T}},
\end{eqnarray}
which can be viewed as the transmit steering vectors in the elevation and azimuth domains. We therefore have $\bm{a}_{\mathrm{t}}(\theta,\phi)=\bm{a}_{\mathrm{tx}}(\mu_{\mathrm{x}})\otimes\bm{a}_{\mathrm{ty}}(\mu_{\mathrm{y}})$. Denote the receive spatial frequency by $\nu=\frac{2\pi}{\lambda}d_{\mathrm{r}}\sin(\psi)$, the receive array response vector is therefore rewritten as
\begin{equation}
\bm{a}_{\mathrm{r}}(\nu)=\frac{1}{\sqrt{M_{\mathrm{tot}}}}\left[1, e^{j\nu},\cdots, e^{j\left(M_{\mathrm{tot}}-1\right)\nu} \right]^{\mathrm{T}}.
\end{equation}

\subsubsection{General setup for analog transmit beamforming and receive combining}
\begin{figure}
\begin{center}
\subfigure[]{%
\includegraphics[width=2.05in]{FD-MIMO-el-az_modify_el.pdf}
\label{fig:subfigure1}}
\quad
\subfigure[]{%
\includegraphics[width=2.5in]{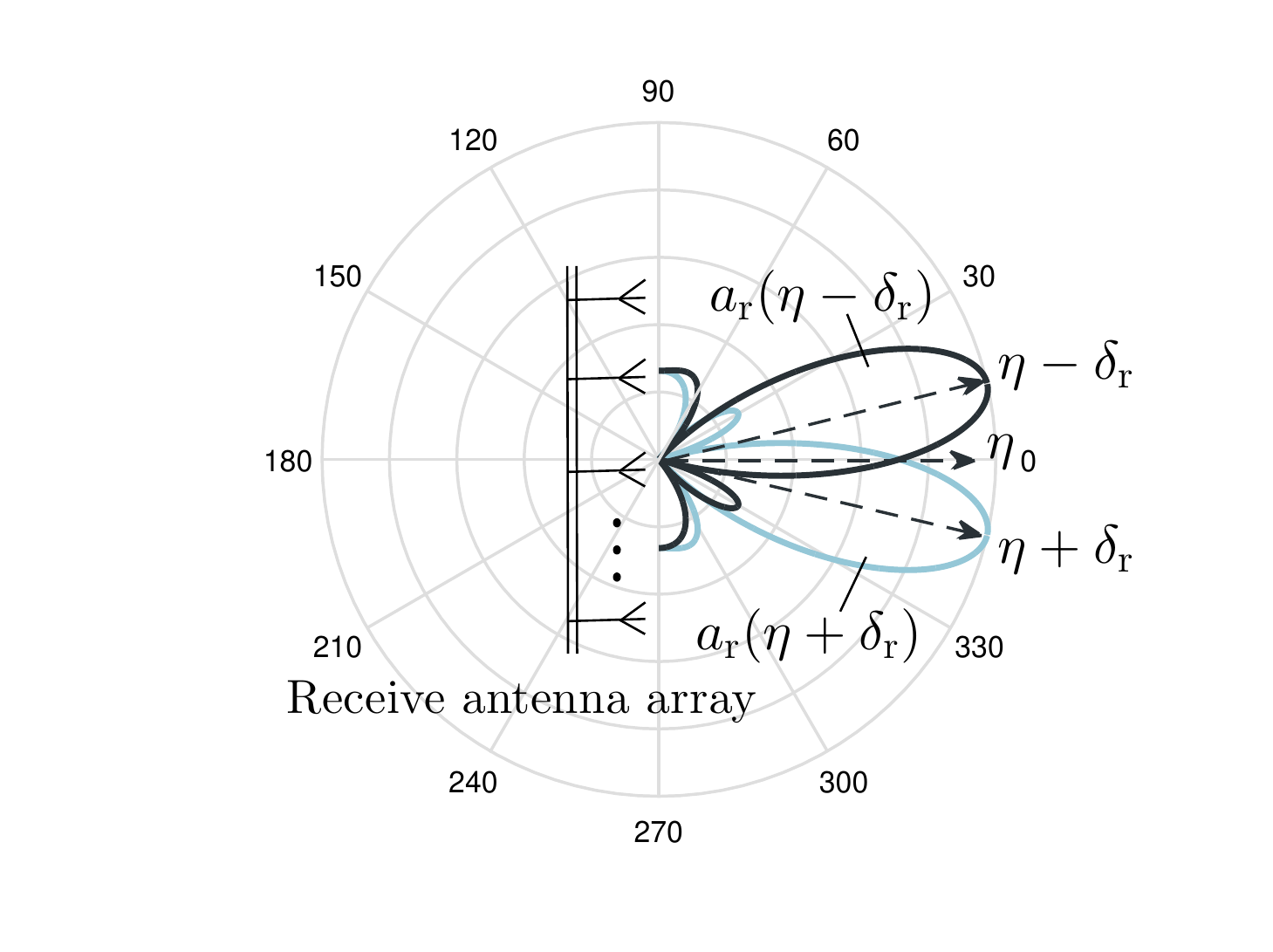}
\label{fig:subfigure1}}
\caption{(a) An example of elevation transmit auxiliary beam pair design. (b) An example of receive beam pair design.}
\label{fig:figure}
\end{center}
\end{figure}
In the proposed auxiliary beam pair based narrowband single-path angle estimation method, both the BS and UE conduct analog-only beamforming and combining using a single RF chain. Pairs of custom designed analog transmit and receive beams are formed to cover the given angular ranges. Each auxiliary beam pair contains two successively probed analog beams in both time and angular domains. In Fig.~2(a), two analog transmit beams form an elevation transmit auxiliary beam pair, and they steer towards $\mu_{\mathrm{el}}-\delta_{\mathrm{x}}$ and $\mu_{\mathrm{el}}+\delta_{\mathrm{x}}$ in the elevation domain for a given azimuth transmit direction $\mu_{\mathrm{az}}'$, i.e., the two transmit array response vectors become
\begin{eqnarray}
\bm{a}_{\mathrm{t}}(\mu_{\mathrm{el}}-\delta_{\mathrm{x}},\mu_{\mathrm{az}}')=\bm{a}_{\mathrm{tx}}(\mu_{\mathrm{el}}-\delta_{\mathrm{x}})\otimes\bm{a}_{\mathrm{ty}}(\mu_{\mathrm{az}}'),\hspace{1.8mm}\bm{a}_{\mathrm{t}}(\mu_{\mathrm{el}}+\delta_{\mathrm{x}},\mu_{\mathrm{az}}')=\bm{a}_{\mathrm{tx}}(\mu_{\mathrm{el}}+\delta_{\mathrm{x}})\otimes\bm{a}_{\mathrm{ty}}(\mu_{\mathrm{az}}').
\end{eqnarray}
Here, $\delta_{\mathrm{x}}=\frac{2\ell_{\mathrm{x}}\pi}{N_{\mathrm{x}}}$ where $\ell_{\mathrm{x}}=1,\cdots,\frac{N_{\mathrm{x}}}{4}$. To form an azimuth transmit auxiliary beam pair, two analog transmit beams are probed, and they steer towards $\mu_{\mathrm{az}}-\delta_{\mathrm{y}}$ and $\mu_{\mathrm{az}}+\delta_{\mathrm{y}}$ in the azimuth domain for a given elevation transmit direction $\mu_{\mathrm{el}}'$. Specifically,
\begin{eqnarray}
\bm{a}_{\mathrm{t}}(\mu_{\mathrm{el}}',\mu_{\mathrm{az}}-\delta_{\mathrm{y}})=\bm{a}_{\mathrm{tx}}(\mu_{\mathrm{el}}')\otimes\bm{a}_{\mathrm{ty}}(\mu_{\mathrm{az}}-\delta_{\mathrm{y}}),\hspace{1.8mm}\bm{a}_{\mathrm{t}}(\mu_{\mathrm{el}}',\mu_{\mathrm{az}}+\delta_{\mathrm{y}})=\bm{a}_{\mathrm{tx}}(\mu_{\mathrm{el}}')\otimes\bm{a}_{\mathrm{ty}}(\mu_{\mathrm{az}}+\delta_{\mathrm{y}}).
\end{eqnarray}
Here, $\delta_{\mathrm{y}}=\frac{2\ell_{\mathrm{y}}\pi}{N_{\mathrm{y}}}$ where $\ell_{\mathrm{y}}=1,\cdots,\frac{N_{\mathrm{y}}}{4}$. Fig.~2(b) shows an example of one receive auxiliary beam pair formed by the UE. The two analog combining vectors targeted at the directions of $\eta-\delta_{\mathrm{r}}$ and $\eta+\delta_{\mathrm{r}}$ are $\bm{a}_{\mathrm{r}}(\eta-\delta_{\mathrm{r}})$ and $\bm{a}_{\mathrm{r}}(\eta+\delta_{\mathrm{r}})$, where $\delta_{\mathrm{r}}=\frac{2\ell_{\mathrm{r}}\pi}{M_{\mathrm{tot}}}$ and $\ell_{\mathrm{r}}=1,\cdots,\frac{M_{\mathrm{tot}}}{4}$.
\subsubsection{Auxiliary beam pair enabled azimuth/elevation AoD and AoA estimation}
We first demonstrate the estimation of azimuth/elevation AoD. Using the analog transmit beam $\bm{a}_{\mathrm{t}}(\mu_{\mathrm{el}}-\delta_{\mathrm{x}},\mu_{\mathrm{az}}')$ and the receive beam $\bm{a}_{\mathrm{r}}(\eta-\delta_{\mathrm{r}})$, the received signal is expressed as
\begin{eqnarray}\label{aoae}
y=g\bm{a}^{*}_{\mathrm{r}}(\eta-\delta_{\mathrm{r}})\bm{a}_{\mathrm{r}}(\nu)\bm{a}^{*}_{\mathrm{t}}(\theta,\phi)\bm{a}_{\mathrm{t}}(\mu_{\mathrm{el}}-\delta_{\mathrm{x}},\mu_{\mathrm{az}}')s_{1}+\bm{a}^{*}_{\mathrm{r}}(\eta-\delta_{\mathrm{r}})\bm{n}.
\end{eqnarray}
Neglecting the noise impairment and assuming $|s_{1}|^2=1$ because of the single RF assumption, the corresponding received signal strength is calculated as
\begin{eqnarray}
\chi_{\mathrm{el}}^{\Delta}&=&|g|^{2}\left|\bm{a}^{*}_{\mathrm{r}}(\eta-\delta_{\mathrm{r}})\bm{a}_{\mathrm{r}}(\nu)\bm{a}^{*}_{\mathrm{t}}(\theta,\phi)\bm{a}_{\mathrm{t}}(\mu_{\mathrm{el}}-\delta_{\mathrm{x}},\mu_{\mathrm{az}}')\right|^{2}\\
&=&|g|^{2}\left|\bm{a}^{*}_{\mathrm{r}}(\eta-\delta_{\mathrm{r}})\bm{a}_{\mathrm{r}}(\nu)\right|^{2}\left|\frac{\sin\left(N_{\mathrm{x}}\left(\frac{\mu_{\mathrm{x}}-\mu_{\mathrm{el}}}{2}\right)\right)}{N_{\mathrm{x}}\sin\left(\frac{\mu_{\mathrm{x}}-\mu_{\mathrm{el}}+\delta_{\mathrm{x}}}{2}\right)}\right|^{2}\left|\frac{\sin\left(N_{\mathrm{y}}\left(\frac{\mu_{\mathrm{y}}-\mu_{\mathrm{az}}'}{2}\right)\right)}{N_{\mathrm{y}}\sin\left(\frac{\mu_{\mathrm{y}}-\mu_{\mathrm{az}}'}{2}\right)}\right|^{2}.\label{dee}
\end{eqnarray}
Similarly, for $\bm{a}_{\mathrm{t}}(\mu_{\mathrm{el}}+\delta_{\mathrm{x}},\mu_{\mathrm{az}}')$ and $\bm{a}_{\mathrm{r}}\left(\eta-\delta_{\mathrm{r}}\right)$, the corresponding received signal strength is
\begin{eqnarray}
\chi_{\mathrm{el}}^{\Sigma}=|g|^{2}\left|\bm{a}^{*}_{\mathrm{r}}(\eta-\delta_{\mathrm{r}})\bm{a}_{\mathrm{r}}(\nu)\right|^{2}\left|\frac{\sin\left(N_{\mathrm{x}}\left(\frac{\mu_{\mathrm{x}}-\mu_{\mathrm{el}}}{2}\right)\right)}{N_{\mathrm{x}}\sin\left(\frac{\mu_{\mathrm{x}}-\mu_{\mathrm{el}}-\delta_{\mathrm{x}}}{2}\right)}\right|^{2}\left|\frac{\sin\left(N_{\mathrm{y}}\left(\frac{\mu_{\mathrm{y}}-\mu_{\mathrm{az}}'}{2}\right)\right)}{N_{\mathrm{y}}\sin\left(\frac{\mu_{\mathrm{y}}-\mu_{\mathrm{az}}'}{2}\right)}\right|^{2}.
\end{eqnarray}
A ratio metric can be obtained as
\begin{equation}
\zeta_{\mathrm{el}}=\frac{\chi_{\mathrm{el}}^{\Delta}-\chi_{\mathrm{el}}^{\Sigma}}{\chi_{\mathrm{el}}^{\Delta}+\chi_{\mathrm{el}}^{\Sigma}}=\frac{\left|\frac{\sin\left(N_{\mathrm{x}}\left(\frac{\mu_{\mathrm{x}}-\mu_{\mathrm{el}}}{2}\right)\right)}{N_{\mathrm{x}}\sin\left(\frac{\mu_{\mathrm{x}}-\mu_{\mathrm{el}}+\delta_{\mathrm{x}}}{2}\right)}\right|^{2}-\left|\frac{\sin\left(N_{\mathrm{x}}\left(\frac{\mu_{\mathrm{x}}-\mu_{\mathrm{el}}}{2}\right)\right)}{N_{\mathrm{x}}\sin\left(\frac{\mu_{\mathrm{x}}-\mu_{\mathrm{el}}-\delta_{\mathrm{x}}}{2}\right)}\right|^{2}}{\left|\frac{\sin\left(N_{\mathrm{x}}\left(\frac{\mu_{\mathrm{x}}-\mu_{\mathrm{el}}}{2}\right)\right)}{N_{\mathrm{x}}\sin\left(\frac{\mu_{\mathrm{x}}-\mu_{\mathrm{el}}+\delta_{\mathrm{x}}}{2}\right)}\right|^{2}+\left|\frac{\sin\left(N_{\mathrm{x}}\left(\frac{\mu_{\mathrm{x}}-\mu_{\mathrm{el}}}{2}\right)\right)}{N_{\mathrm{x}}\sin\left(\frac{\mu_{\mathrm{x}}-\mu_{\mathrm{el}}-\delta_{\mathrm{x}}}{2}\right)}\right|^{2}}=-\frac{\sin\left(\frac{\mu_{\mathrm{x}}-\mu_{\mathrm{el}}}{2}\right)\sin(\delta_{\mathrm{x}})}{1-\cos\left(\mu_{\mathrm{x}}-\mu_{\mathrm{el}}\right)\cos(\delta_{\mathrm{x}})},\label{diffmono}
\end{equation}
which does not depend the path gain, analog receive combining and azimuth transmit beamforming, in the absence of noise. If $|\mu_{\mathrm{x}}-\mu_{\mathrm{el}}|<\delta_{\mathrm{x}}$, i.e., $\mu_{\mathrm{x}}$ is within the range of $\left(\mu_{\mathrm{el}}-\delta_{\mathrm{x}},\mu_{\mathrm{el}}+\delta_{\mathrm{x}}\right)$, $\zeta_{\mathrm{el}}$ is a monotonically decreasing function of $\mu_{\mathrm{x}}-\mu_{\mathrm{el}}$ and invertible with respect to $\mu_{\mathrm{x}}-\mu_{\mathrm{el}}$ \cite[Lemma~1]{dztrans}. Hence, by solving (\ref{diffmono}), we have
\begin{eqnarray}\label{aoangle}
\hat{\mu}_{\mathrm{x}}=\mu_{\mathrm{el}}-\arcsin\Bigg(\frac{\zeta_{\mathrm{el}}\sin(\delta_{\mathrm{x}})-\zeta_{\mathrm{el}}\sqrt{1-\zeta_{\mathrm{el}}^{2}}\sin(\delta_{\mathrm{x}})\cos(\delta_{\mathrm{x}})}{\sin^{2}(\delta_{\mathrm{x}})+\zeta_{\mathrm{el}}^{2}\cos^{2}(\delta_{\mathrm{x}})}\Bigg).
\end{eqnarray}
Note that if $\zeta_{\mathrm{el}}$ is not impaired by noise, $\mu_{\mathrm{x}}=\hat{\mu}_{\mathrm{x}}$, implying the super-resolution estimation.

Similarly, neglecting noise and assuming $\bm{a}_{\mathrm{r}}(\eta+\delta_{\mathrm{r}})$ is the analog combining vector formed by the UE, for the azimuth transmit auxiliary beam pair containing $\bm{a}_{\mathrm{t}}(\mu_{\mathrm{el}}',\mu_{\mathrm{az}}-\delta_{\mathrm{y}})$ and $\bm{a}_{\mathrm{t}}(\mu_{\mathrm{el}}',\mu_{\mathrm{az}}+\delta_{\mathrm{y}})$, we can obtain
\begin{equation}\label{adhoc}
\zeta_{\mathrm{az}}=-\frac{\sin\left(\frac{\mu_{\mathrm{y}}-\mu_{\mathrm{az}}}{2}\right)\sin(\delta_{\mathrm{y}})}{1-\cos\left(\mu_{\mathrm{y}}-\mu_{\mathrm{az}}\right)\cos(\delta_{\mathrm{y}})},
\end{equation}
and therefore, if $|\mu_{\mathrm{y}}-\mu_{\mathrm{az}}|<\delta_{\mathrm{y}}$,
\begin{eqnarray}\label{elaoangle}
\hat{\mu}_{\mathrm{y}}=\mu_{\mathrm{az}}-\arcsin\Bigg(\frac{\zeta_{\mathrm{az}}\sin(\delta_{\mathrm{y}})-\zeta_{\mathrm{az}}\sqrt{1-\zeta_{\mathrm{az}}^{2}}\sin(\delta_{\mathrm{y}})\cos(\delta_{\mathrm{y}})}{\sin^{2}(\delta_{\mathrm{y}})+\zeta_{\mathrm{az}}^{2}\cos^{2}(\delta_{\mathrm{y}})}\Bigg).
\end{eqnarray}
Using $\hat{\mu}_{\mathrm{x}}$ and $\hat{\mu}_{\mathrm{y}}$, the estimated azimuth and elevation AoDs are obtained as
\begin{equation}\label{bpaz}
\hat{\phi}=\arctan\left(\frac{d_{\mathrm{tx}}\hat{\mu}_{\mathrm{y}}}{d_{\mathrm{ty}}\hat{\mu}_{\mathrm{x}}}\right),\hspace{4mm}\hat{\theta}=\arcsin\left(\frac{\lambda\hat{\mu}_{\mathrm{y}}}{2\pi d_{\mathrm{ty}}}\left(\sin\left(\arctan\left(\frac{d_{\mathrm{tx}}\hat{\mu}_{\mathrm{y}}}{d_{\mathrm{ty}}\hat{\mu}_{\mathrm{x}}}\right)\right)\right)^{-1}\right).
\end{equation}

The estimation of receive spatial frequency and AoA via the receive auxiliary beam pair can be conducted in a similar fashion.
\subsubsection{Practical implementation for the proposed approach}
In practice, the relative positions of the azimuth/elevation AoD and AoA to the auxiliary beam pairs, i.e., the assumptions $|\mu_{\mathrm{x}}-\mu_{\mathrm{el}}|<\delta_{\mathrm{x}}$, $|\mu_{\mathrm{y}}-\mu_{\mathrm{az}}|<\delta_{\mathrm{y}}$ and $|\nu-\eta|<\delta_{\mathrm{r}}$ made in the previous examples, are unknown a prior. Multiple auxiliary beam pairs are therefore formed by both the BS and UE to cover given angular ranges. As single RF chain is employed, the analog transmit and receive beams are probed in a round-robin time division multiplexing (TDM) manner. For instance, for a given analog receive beam, all analog transmit beams covering both the azimuth and elevation domains are successively probed by the BS. This process continues until all analog receive beams have been probed. Denote the total numbers of probed analog transmit beams and receive beams by $N_{\mathrm{TX}}$ and $N_{\mathrm{RX}}$. The total number of iterations between the BS and UE can then be computed as $N_{\mathrm{TX}}N_{\mathrm{RX}}$.

Further, as the noise impairment is neglected during the derivations, the calculated ratio metric for estimating the azimuth (elevation) transmit spatial frequency becomes independent of the receive combining vector and the elevation (azimuth) steering direction. To implement the proposed algorithm in practical cellular systems with noise impairment and various types of interferences, it is critical to select the best radio link between the transmit (or receive) auxiliary beam pair and the corresponding receive (or transmit) beam among all $N_{\mathrm{TX}}N_{\mathrm{RX}}$ combinations. According to \cite[Lemma~2]{dztrans}, this can be done by simple received signal power comparison such that the selected transmit/receive auxiliary beam pair covers the AoD/AoA to be estimated with high probability. Similar design principle is adopted in Section III-B to choose the best combinations between transmit (receive) auxiliary beam pairs and receive (transmit) beams for multi-path angle estimation.

\subsection{Wideband multi-path angle estimation using multiple RF chains with cross-polarization}
\begin{figure}
\centering
\includegraphics[width=3.2in]{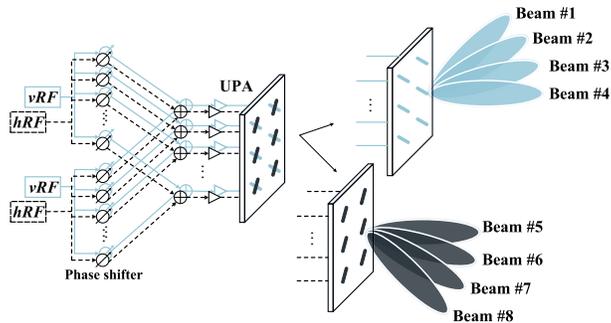}
\caption{A conceptual example of the mapping between the polarization domain and auxiliary beam pair. In this example, every two beams with consecutive indices probed from the same polarization domain form an auxiliary beam pair. For instance, Beams $\#1$ and $\#2$ form an auxiliary beam pair, while Beams $\#1$ and $\#4$ or Beams $\#4$ and $\#5$ are not paired.}
\end{figure}
To facilitate the auxiliary beam pair enabled multi-path angle estimation for wideband channels, multiple RF chains can be employed by both the BS and UE such that multiple auxiliary beam pairs can be probed simultaneously. To distinguish between simultaneously probed auxiliary beam pairs, a first-layer auxiliary beam pair specific pilot design is developed.

In the proposed design approach, the two beams in one auxiliary beam pair are generated from the same polarization domain. A conceptual example illustrating this is given in Fig.~3, in which the beams formed from the vertically polarized antennas are indexed from $1\sim 4$, and the beams probed from the horizontal domain are indexed from $5\sim 8$. Denote the vertical and horizontal beam index sets by $\Upsilon_{\mathrm{v}}=\left\{\#1, \#2, \#3, \#4\right\}$ and $\Upsilon_{\mathrm{h}}=\left\{\#5, \#6, \#7, \#8\right\}$. Only the beams that are chosen from the same set (either $\Upsilon_{\mathrm{v}}$ or $\Upsilon_{\mathrm{h}}$) can be paired. For instance, the two beams with beam indices $\#4$ and $\#5$ selected from $\Upsilon_{\mathrm{v}}$ and $\Upsilon_{\mathrm{h}}$ are not allowed to form an auxiliary beam pair. As will be discussed later, by pairing beams only from the same polarization domain, the corresponding ratio metrics derived assuming high-power regime will exhibit similar forms to those using co-polarized antennas, i.e., (\ref{diffmono}) and (\ref{adhoc}).

A second-layer paired-beam specific pilot is therefore developed and applied to distinguish between the two beams in the same auxiliary beam pair. In this paper, the Zadoff-Chu (ZC) sequences are used to design the proposed multi-layer pilot. In the current 3GPP LTE standards, the ZC sequences are used as primary synchronization signals (PSS) for cell search, sounding reference signal (RS) for uplink channel estimation, demodulation RS for uplink control and data channels, and random access channel preambles \cite{lte}. The benefits of applying the multi-layer pilot structure to the proposed approach is elaborated at the end of this section after the detailed algorithm design has been illustrated.

For illustration, the estimation of azimuth transmit spatial frequency is of our design focus throughout this part, while the estimation of elevation transmit spatial frequency and receive spatial frequency can be derived in a similar fashion.
\subsubsection{General setup for analog transmit beamforming and receive combining}
For a given elevation transmit direction $\mu_{\mathrm{el}}'$, denote by $\mathcal{F}_{\mathrm{T}}^{\mathrm{v}}$ and $\mathcal{F}_{\mathrm{T}}^{\mathrm{h}}$ the codebooks of azimuth analog transmit steering vectors for the vertical and horizontal transmit RF chains, we have
\begin{equation}\label{bmfs}
\mathcal{F}_{\mathrm{T}}^{\mathrm{v}}=\left\{\left[\bm{a}^{\mathrm{T}}_{\mathrm{t}}\left(\mu_{\mathrm{el}}',\mu_{\mathrm{az}}^{0}\right)\hspace{2mm}\bm{0}^{\mathrm{T}}_{N_{\mathrm{tot}}/2}\right]^{\mathrm{T}},\cdots,\left[\bm{a}^{\mathrm{T}}_{\mathrm{t}}\left(\mu_{\mathrm{el}}',\mu_{\mathrm{az}}^{N_{\mathrm{az}}/2-1}\right)\hspace{2mm}\bm{0}^{\mathrm{T}}_{N_{\mathrm{tot}}/2}\right]^{\mathrm{T}}\right\},
\end{equation}
\begin{equation}
\mathcal{F}_{\mathrm{T}}^{\mathrm{h}}=\left\{\left[\bm{0}^{\mathrm{T}}_{N_{\mathrm{tot}}/2}\hspace{2mm}\bm{a}^{\mathrm{T}}_{\mathrm{t}}\left(\mu_{\mathrm{el}}',\mu_{\mathrm{az}}^{N_{\mathrm{az}}/2}\right)\right]^{\mathrm{T}},\cdots,\left[\bm{0}^{\mathrm{T}}_{N_{\mathrm{tot}}/2}\hspace{2mm}\bm{a}^{\mathrm{T}}_{\mathrm{t}}\left(\mu_{\mathrm{el}}',\mu_{\mathrm{az}}^{N_{\mathrm{az}}-1}\right)\right]^{\mathrm{T}}\right\},
\end{equation}
where $\left\{\mu_{\mathrm{az}}^{0},\cdots,\mu_{\mathrm{az}}^{N_{\mathrm{az}}-1}\right\}$ is the set of candidate azimuth transmit directions. Similarly,
\begin{equation}
\mathcal{W}_{\mathrm{T}}^{\mathrm{v}}=\left\{\left[\bm{a}^{\mathrm{T}}_{\mathrm{r}}\left(\nu^{0}\right)\hspace{2mm}\bm{0}^{\mathrm{T}}_{M_{\mathrm{tot}}/2}\right]^{\mathrm{T}},\cdots,\left[\bm{a}^{\mathrm{T}}_{\mathrm{r}}\left(\nu^{N_{\mathrm{rx}}/2-1}\right)\hspace{2mm}\bm{0}^{\mathrm{T}}_{M_{\mathrm{tot}}/2}\right]^{\mathrm{T}}\right\},
\end{equation}
\begin{equation}\label{bmwe}
\mathcal{W}_{\mathrm{T}}^{\mathrm{h}}=\left\{\left[\bm{0}^{\mathrm{T}}_{M_{\mathrm{tot}}/2}\hspace{2mm}\bm{a}^{\mathrm{T}}_{\mathrm{r}}\left(\nu^{N_{\mathrm{rx}}/2}\right)\right]^{\mathrm{T}},\cdots,\left[\bm{0}^{\mathrm{T}}_{M_{\mathrm{tot}}/2}\hspace{2mm}\bm{a}^{\mathrm{T}}_{\mathrm{r}}\left(\nu^{N_{\mathrm{rx}}-1}\right)\right]^{\mathrm{T}}\right\},
\end{equation}
denote the codebooks of analog receive steering vectors for the vertical and horizontal receive RF chains, where $\left\{\nu^{0},\cdots,\nu^{N_{\mathrm{rx}}-1}\right\}$ is the set of candidate receive directions. Note that here, $\bm{a}_{\mathrm{t}}\left(\mu_{\mathrm{el}}',\mu_{\mathrm{az}}^{n}\right)\in\mathbb{C}^{\frac{N_{\mathrm{tot}}}{2}\times 1}$ and $\bm{a}_{\mathrm{r}}\left(\nu^{\acute{n}}\right)\in\mathbb{C}^{\frac{M_{\mathrm{tot}}}{2}\times 1}$ due to the cross-polarization assumption, where $n\in\left\{0,\cdots,N_{\mathrm{az}}-1\right\}$ and $\acute{n}\in\left\{0,\cdots,N_{\mathrm{rx}}-1\right\}$.

The constructions of the transmit and receive beam codebooks in (\ref{bmfs})-(\ref{bmwe}) leverage the fact that the AoD and AoA are the same for both vertical and horizontal polarization domains. It is therefore feasible to form vertical beams to cover one half of the probing range, and horizontal beams to cover the other half. By doing so, the number of required beams to cover a given angular range is minimized, which in turn, reduces the estimation overhead.

The analog transmit and receive probing matrices are constructed by concatenating all successively probed azimuth analog transmit precoding and receive combining matrices. Denote by $N_{\mathrm{T}}$ and $M_{\mathrm{T}}$ the total numbers of probings performed by the BS in the azimuth domain for a given elevation steering direction and UE. Further, denote by $\bm{F}_{\mathrm{T}}$ and $\bm{W}_{\mathrm{T}}$ the analog azimuth transmit and receive probing matrices, we have $\bm{F}_{\mathrm{T}}=\left[\bm{F}_{1},\cdots,\bm{F}_{n_{\mathrm{t}}},\cdots,\bm{F}_{N_{\mathrm{T}}}\right]$ and $\bm{W}_{\mathrm{T}}=\left[\bm{W}_{1},\cdots,\bm{W}_{m_{\mathrm{t}}},\cdots,\bm{W}_{M_{\mathrm{T}}}\right]$, where $\bm{F}_{n_{\mathrm{t}}}\in\mathbb{C}^{N_{\mathrm{tot}}\times N_{\mathrm{RF}}}$ represents the $n_{\mathrm{t}}$-th azimuth transmit probing, $\bm{W}_{m_{\mathrm{t}}}\in\mathbb{C}^{M_{\mathrm{tot}}\times M_{\mathrm{RF}}}$ is the $m_{\mathrm{t}}$-th receive probing.

To facilitate the selection of the best pairs of azimuth transmit auxiliary beam pairs and receive beams with noise impairment and various interference sources, the steering angles of simultaneously probed beams should match the distribution of azimuth AoD and AoA, which are unknown to the BS and UE a prior, as much as possible. With finite $N_{\mathrm{T}}$ and $M_{\mathrm{T}}$, the analog beams in one probing matrix are steered towards random angular directions. That is, each column in $\bm{F}_{n_{\mathrm{t}}}$ is randomly chosen from $\mathcal{F}_{\mathrm{T}}^{\mathrm{v}}$ and $\mathcal{F}_{\mathrm{T}}^{\mathrm{h}}$, and each column in $\bm{W}_{m_{\mathrm{t}}}$ is randomly chosen from $\mathcal{W}_{\mathrm{T}}^{\mathrm{v}}$ and $\mathcal{W}_{\mathrm{T}}^{\mathrm{h}}$. Note that from (\ref{bmfs})-(\ref{bmwe}), it can be observed that the vertical and horizontal beams cover the first and second half of the probing range, the transmit and receive beams can therefore be randomly selected from the same polarization domain for a given probing to facilitate the whole process. The transmit and receive probings are conducted in a TDM manner. For a given probing at the UE, $N_{\mathrm{T}}$ consecutive probings $\bm{F}_{1},\cdots,\bm{F}_{N_{\mathrm{T}}}$ are performed at the BS. This process iterates until all $M_{\mathrm{T}}$ probings have been executed by the UE.

Consider a given probing at the UE, e.g., $\bm{W}_{m_{\mathrm{t}}}$, the resultant matrix by concatenating the $N_{\mathrm{T}}N_{\mathrm{RF}}$ azimuth transmit beamforming vectors in the absence of noise is obtained as
\begin{eqnarray}
\bm{Y}_{m_{\mathrm{t}}}[k]&=&\bm{W}_{m_{\mathrm{t}}}^{*}\bm{H}[k]\bm{F}_{\mathrm{T}}\bm{X}[k],\label{sp}
\end{eqnarray}
where $\bm{X}[k]$ carries the $N_{\mathrm{T}}N_{\mathrm{RF}}$ multi-layer pilot symbols such that
\begin{equation}\label{sprs}
\left[\bm{X}[k]\right]_{i,q}=\bigg\{\begin{array}{l}
                                    0, \hspace{2mm} 0<i\leq N_{\mathrm{RF}}(q-1)\hspace{1.5mm} \textrm{or} \hspace{1.5mm}N_{\mathrm{RF}}q<i\leq N_{\mathrm{T}}N_{\mathrm{RF}}, \hspace{1.5mm} q=1,\cdots,N_{\mathrm{T}}, \\
\neq0, \hspace{2mm}\textrm{otherwise}.
                                  \end{array}
\end{equation}


\subsubsection{Design principles and procedures for the proposed approach}

The ZC sequences are used for designing the multi-layer pilot due to their constant amplitude and zero autocorrelation properties in both time and frequency domains \cite{popov}. In this paper, the multi-layer pilot consists of a first-layer auxiliary beam pair specific pilot and a second-layer paired-beam specific pilot. Denote by $\mathrm{a}_{\ell}$ and $\mathrm{b}_{m}$ the auxiliary beam pair identity (ID) and paired-beam ID. Denote by $N_{\mathrm{a}}$ the total number of possible azimuth transmit auxiliary beam pairs that can be constructed via the azimuth transmit steering vectors in $\mathcal{F}_{\mathrm{T}}^{\mathrm{v}}$ and $\mathcal{F}_{\mathrm{T}}^{\mathrm{h}}$, we have $\ell\in\left\{0,\cdots,N_{\mathrm{a}}-1\right\}$ and $m\in\{0,\cdots,N_{\mathrm{az}}-1\}$. Note that as one auxiliary beam pair consists of two beams, $\mathrm{b}_{m}\in\left\{0,1\right\}$. By mapping the proposed multi-layer pilot to one OFDM symbol, we have
\begin{equation}\label{map1}
x^{\mathrm{a}_{\ell},\mathrm{b}_{m}}[k]=e^{\frac{j\pi r^{(\mathrm{a}_{\ell})}(k+p\mathrm{b}_{m})(k+p\mathrm{b}_{m}+1)}{N}},
\end{equation}
where $r^{(\mathrm{a}_{\ell})}$ denotes the root index associated with the auxiliary beam pair ID $\mathrm{a}_{\ell}$, and $p$ is a prime number representing the circular shift spacing in the frequency domain. Further, $\mathrm{gcd}(r^{(\mathrm{a}_{\ell})},N)=1$ is satisfied. Assume that the azimuth transmit beamforming vector $\left[\bm{F}_{\mathrm{T}}\right]_{:,q}$ corresponds to the auxiliary beam pair ID $\mathrm{a}_{\ell}$ and the paired-beam ID $\mathrm{b}_{m}$, we have $\left[\bm{X}[k]\right]_{q,\lceil q/N_{\mathrm{RF}}\rceil}=x^{\mathrm{a}_{\ell},\mathrm{b}_{m}}[k]$. Note that as can be seen from (\ref{map1}), we reuse the ZC sequence structure as in \cite{lte}, but with custom designed mapping between the root indices, frequency domain circular shifts and auxiliary beams.

Consider the $n_{\mathrm{t}}$-th azimuth probing $\bm{F}_{n_{\mathrm{t}}}$ at the BS. For simplicity, the following assumptions are made: (i) $N_{\mathrm{r}}=M_{\mathrm{RF}}$, and (ii) the first $N_{\mathrm{RF}}/2$ columns in $\bm{F}_{n_{\mathrm{t}}}$ are chosen from $\mathcal{F}_{\mathrm{T}}^{\mathrm{v}}$, while its last $N_{\mathrm{RF}}/2$ columns are selected from $\mathcal{F}_{\mathrm{T}}^{\mathrm{h}}$ assuming that $N_{\mathrm{RF}}$ is an even number. Specifically,
\begin{align}\label{specase}
\bm{F}_{n_{\mathrm{t}}}=\Big[&\left[\bm{a}^{\mathrm{T}}_{\mathrm{t}}\left(\mu_{\mathrm{el}}',\mu_{\mathrm{az}}^{n_{0}}\right)\hspace{2mm}\bm{0}^{\mathrm{T}}_{N_{\mathrm{tot}}/2}\right]^{\mathrm{T}}\hspace{2mm}\cdots\hspace{2mm}\left[\bm{a}^{\mathrm{T}}_{\mathrm{t}}\left(\mu_{\mathrm{el}}',\mu_{\mathrm{az}}^{n_{N_{\mathrm{RF}}/2-1}}\right)\hspace{2mm}\bm{0}^{\mathrm{T}}_{N_{\mathrm{tot}}/2}\right]^{\mathrm{T}}\nonumber\\
&\left[\bm{0}^{\mathrm{T}}_{N_{\mathrm{tot}}/2}\hspace{2mm}\bm{a}^{\mathrm{T}}_{\mathrm{t}}\left(\mu_{\mathrm{el}}',\mu_{\mathrm{az}}^{n_{N_{\mathrm{RF}}/2}}\right)\right]^{\mathrm{T}}\hspace{2mm}\cdots\hspace{2mm}\left[\bm{0}^{\mathrm{T}}_{N_{\mathrm{tot}}/2}\hspace{2mm}\bm{a}^{\mathrm{T}}_{\mathrm{t}}\left(\mu_{\mathrm{el}}',\mu_{\mathrm{az}}^{n_{N_{\mathrm{RF}}-1}}\right)\right]^{\mathrm{T}}\Big],
\end{align}
where $n_{0},\cdots,n_{N_{\mathrm{RF}}/2-1}\in\left\{0,\cdots,N_{\mathrm{az}}/2-1\right\}$ and $n_{N_{\mathrm{RF}}/2},\cdots,n_{N_{\mathrm{RF}}-1}\in\left\{N_{\mathrm{az}}/2,\cdots,N_{\mathrm{az}}-1\right\}$. Further, assume that the azimuth transmit beamforming vectors $\bm{a}_{\mathrm{t}}\left(\mu_{\mathrm{el}}',\mu_{\mathrm{az}}^{n_{0}}\right),\cdots,\bm{a}_{\mathrm{t}}\left(\mu_{\mathrm{el}}',\mu_{\mathrm{az}}^{n_{N_{\mathrm{RF}}-1}}\right)$ correspond to the auxiliary beam pair IDs $\mathrm{a}_{0},\cdots,\mathrm{a}_{N_{\mathrm{RF}}-1}$ and the paired-beam IDs $\mathrm{b}_{0},\cdots,\mathrm{b}_{N_{\mathrm{RF}}-1}$.

Before proceeding with detailed design procedures, we first partition the channel model in (\ref{wbfrp}) into a block matrix form as
\begin{equation}
\bm{H}[k]=\left[
            \begin{array}{cc}
              \bm{H}^{\mathrm{vv}}[k] & \bm{H}^{\mathrm{vh}}[k] \\
              \bm{H}^{\mathrm{hv}}[k] & \bm{H}^{\mathrm{hh}}[k] \\
            \end{array}
          \right],\hspace{2mm}\bm{H}^{\mathrm{vv}}[k],\bm{H}^{\mathrm{vh}}[k],\bm{H}^{\mathrm{hv}}[k],\bm{H}^{\mathrm{vv}}[k]\in\mathbb{C}^{\frac{M_{\mathrm{tot}}}{2}\times\frac{N_{\mathrm{tot}}}{2}}.
\end{equation}
Denote by $\bm{A}_{r}^{\mathrm{TR}}=\bm{1}_{\frac{M_{\mathrm{tot}}}{2}\times\frac{N_{\mathrm{tot}}}{2}}\odot\left(\bm{a}_{\mathrm{r}}\left(\psi_{r}\right)\bm{a}_{\mathrm{t}}^{*}\left(\theta_r,\phi_r\right)\right)$, we have
\begin{eqnarray}
\bm{H}^{\mathrm{vv}}[k]&=&\sqrt{\frac{1}{1+\chi}}\sum_{r=1}^{N_{\mathrm{r}}}\rho_{\tau_r}[k]\left(g_{r}^{\mathrm{vv}}\cos(\varsigma)+\sqrt{\chi}g_{r}^{\mathrm{vh}}\sin(\varsigma)\right)\bm{A}_{r}^{\mathrm{TR}},\\
\bm{H}^{\mathrm{vh}}[k]&=&\sqrt{\frac{1}{1+\chi}}\sum_{r=1}^{N_{\mathrm{r}}}\rho_{\tau_r}[k]\left(-g_{r}^{\mathrm{vv}}\sin(\varsigma)+\sqrt{\chi}g_{r}^{\mathrm{vh}}\cos(\varsigma)\right)\bm{A}_{r}^{\mathrm{TR}},\\
\bm{H}^{\mathrm{hv}}[k]&=&\sqrt{\frac{1}{1+\chi}}\sum_{r=1}^{N_{\mathrm{r}}}\rho_{\tau_r}[k]\left(\sqrt{\chi}g_{r}^{\mathrm{hv}}\cos(\varsigma)+g_{r}^{\mathrm{hh}}\sin(\varsigma)\right)\bm{A}_{r}^{\mathrm{TR}},\\
\bm{H}^{\mathrm{hh}}[k]&=&\sqrt{\frac{1}{1+\chi}}\sum_{r=1}^{N_{\mathrm{r}}}\rho_{\tau_r}[k]\left(-\sqrt{\chi}g_{r}^{\mathrm{hv}}\sin(\varsigma)+g_{r}^{\mathrm{hh}}\cos(\varsigma)\right)\bm{A}_{r}^{\mathrm{TR}}.
\end{eqnarray}

In the absence of noise and consider the $r$-th receive beam $\left[\bm{W}_{m_{\mathrm{t}}}\right]_{:,r}$ chosen from $\mathcal{W}_{\mathrm{T}}^{\mathrm{v}}$,
\begin{eqnarray}
\left[\bm{Y}_{m_{\mathrm{t}}}[k]\right]_{r,n_{\mathrm{t}}}&=&\sum_{\imath=0}^{N_{\mathrm{RF}}/2-1}\bm{w}^{*}_{m_{\mathrm{t}},r}\bm{H}^{\mathrm{vv}}[k]\bm{a}_{\mathrm{t}}\left(\mu_{\mathrm{el}}',\mu_{\mathrm{az}}^{n_{\imath}}\right)x^{\mathrm{a}_{\imath},\mathrm{b}_{\imath}}[k]\nonumber\\
&+&\sum_{\kappa=N_{\mathrm{RF}}/2}^{N_{\mathrm{RF}}-1}\bm{w}^{*}_{m_{\mathrm{t}},r}\bm{H}^{\mathrm{vh}}[k]\bm{a}_{\mathrm{t}}\left(\mu_{\mathrm{el}}',\mu_{\mathrm{az}}^{n_{\kappa}}\right)x^{\mathrm{a}_{\kappa},\mathrm{b}_{\kappa}}[k],
\end{eqnarray}
where $\bm{w}_{m_{\mathrm{t}},r}=\left[\bm{W}_{m_{\mathrm{t}}}\right]_{1:\frac{M_{\mathrm{tot}}}{2},r}$. For $\imath\in\left\{0,\cdots,N_{\mathrm{RF}}/2-1\right\}$, we have
\begin{align}
&\bm{w}^{*}_{m_{\mathrm{t}},r}\bm{H}^{\mathrm{vv}}[k]\bm{a}_{\mathrm{t}}\left(\mu_{\mathrm{el}}',\mu_{\mathrm{az}}^{n_{\imath}}\right)\nonumber\\
=&\sqrt{\frac{1}{1+\chi}}\sum_{r'=1}^{N_{\mathrm{r}}}\rho_{\tau_{r'}}[k]\left(g_{r'}^{\mathrm{vv}}\cos(\varsigma)+\sqrt{\chi}g_{r'}^{\mathrm{vh}}\sin(\varsigma)\right)\nonumber\\
&\times\left(\bm{w}^{*}_{m_{\mathrm{t}},r}\left(\bm{1}_{\frac{M_{\mathrm{tot}}}{2}\times\frac{N_{\mathrm{tot}}}{2}}\odot\left(\bm{a}_{\mathrm{r}}\left(\psi_{r'}\right)\right)\bm{a}_{\mathrm{t}}^{*}\left(\theta_{r'},\phi_{r'}\right)\right)\bm{a}_{\mathrm{t}}\left(\mu_{\mathrm{el}}',\mu_{\mathrm{az}}^{n_{\imath}}\right)\right)\\
=&\bm{w}^{*}_{m_{\mathrm{t}},r}\sum_{r'=1}^{N_{\mathrm{r}}}\rho_{\tau_{r'}}[k]h_{r'}^{\mathrm{vv}}\bm{a}_{\mathrm{r}}\left(\psi_{r'}\right)\bm{a}_{\mathrm{t}}^{*}\left(\theta_{r'},\phi_{r'}\right)\bm{a}_{\mathrm{t}}\left(\mu_{\mathrm{el}}',\mu_{\mathrm{az}}^{n_{\imath}}\right),\label{flpf}
\end{align}
where $h_{r'}^{\mathrm{vv}}=\sqrt{\frac{1}{1+\chi}}\left(g_{r'}^{\mathrm{vv}}\cos(\varsigma)+\sqrt{\chi}g_{r'}^{\mathrm{vh}}\sin(\varsigma)\right)$. Similarly, for $\kappa\in\left\{N_{\mathrm{RF}}/2,\cdots,N_{\mathrm{RF}}-1\right\}$,
\begin{equation}
\bm{w}^{*}_{m_{\mathrm{t}},r}\bm{H}^{\mathrm{vh}}[k]\bm{a}_{\mathrm{t}}\left(\mu_{\mathrm{el}}',\mu_{\mathrm{az}}^{n_{\kappa}}\right)=\bm{w}^{*}_{m_{\mathrm{t}},r}\sum_{r'=1}^{N_{\mathrm{r}}}\rho_{\tau_{r'}}[k]h^{\mathrm{vh}}_{r'}\bm{a}_{\mathrm{r}}\left(\psi_{r'}\right)\bm{a}_{\mathrm{t}}^{*}\left(\theta_{r'},\phi_{r'}\right)\bm{a}_{\mathrm{t}}\left(\mu_{\mathrm{el}}',\mu_{\mathrm{az}}^{n_{\kappa}}\right),
\end{equation}
where $h_{r'}^{\mathrm{vh}}=\sqrt{\frac{1}{1+\chi}}\left(-g_{r'}^{\mathrm{vv}}\sin(\varsigma)+\sqrt{\chi}g_{r'}^{\mathrm{vh}}\cos(\varsigma)\right)$.

At the UE side, the received signal in the frequency domain is correlated with the reference ZC sequence corresponding to the reference auxiliary beam pair ID $\mathrm{a}_{\bar{\ell}}$ and the reference paired-beam ID $\mathrm{b}_{\bar{m}}$ ($\bar{\ell}\in\left\{0,\cdots,N_{\mathrm{a}}-1\right\}$, $\bar{m}\in\left\{0,\cdots,N_{\mathrm{az}}-1\right\}$) at zero-lag. In the following, we will interpret the interference terms from the perspective of the vertical polarization domain, while the same principles can be applied to the horizontal polarization domain as well. Specifically,
\begin{align}
&y^{u}_{m_{\mathrm{t}},r,n_{\mathrm{t}}}=\sum_{k=0}^{N-1}\left[\bm{Y}_{m_{\mathrm{t}}}[k]\right]_{r,n_{\mathrm{t}}}\left(x^{\mathrm{a}_{\bar{\ell}},\mathrm{b}_{\bar{m}}}[k]\right)^{*}\\
&=\underbrace{\sum_{k=0}^{N-1}\bm{w}^{*}_{m_{\mathrm{t}},r}\left(\sum_{r'=1}^{N_{\mathrm{r}}}\rho_{\tau_{r'}}[k]h_{r'}^{\mathrm{vv}}\bm{a}_{\mathrm{r}}(\psi_{r'})\bm{a}_{\mathrm{t}}^{*}(\theta_{r'},\phi_{r'})\right)\bm{a}_{\mathrm{t}}\left(\mu_{\mathrm{el}}',\mu_{\mathrm{az}}^{n_{u}}\right)x^{\mathrm{a}_{u},\mathrm{b}_{u}}[k]\left(x^{\mathrm{a}_{\bar{\ell}},\mathrm{b}_{\bar{m}}}[k]\right)^{*}}_{I_{0}}\label{i0i1i2}\\
&+\underbrace{\sum_{k=0}^{N-1}\bm{w}^{*}_{m_{\mathrm{t}},r}\left(\sum_{r'=1}^{N_{\mathrm{r}}}\rho_{\tau_{r'}}[k]h_{r'}^{\mathrm{vv}}\bm{a}_{\mathrm{r}}(\psi_{r'})\bm{a}_{\mathrm{t}}^{*}(\theta_{r'},\phi_{r'})\right)\bm{a}_{\mathrm{t}}\left(\mu_{\mathrm{el}}',\mu_{\mathrm{az}}^{n_{\bar{u}}}\right)x^{\mathrm{a}_{\bar{u}},\mathrm{b}_{\bar{u}}}[k]\left(x^{\mathrm{a}_{\bar{\ell}},\mathrm{b}_{\bar{m}}}[k]\right)^{*}}_{I_{1}}\nonumber\\
&+\underbrace{\sum_{k=0}^{N-1}\sum_{\breve{u}\neq\bar{\ell}}\bm{w}^{*}_{m_{\mathrm{t}},r}\left(\sum_{r'=1}^{N_{\mathrm{r}}}\rho_{\tau_{r'}}[k]h_{r'}^{\mathrm{vv}}\bm{a}_{\mathrm{r}}(\psi_{r'})\bm{a}_{\mathrm{t}}^{*}(\theta_{r'},\phi_{r'})\right)\bm{a}_{\mathrm{t}}\left(\mu_{\mathrm{el}}',\mu_{\mathrm{az}}^{n_{\breve{u}}}\right)x^{\mathrm{a}_{\breve{u}},\mathrm{b}_{\breve{u}}}[k]\left(x^{\mathrm{a}_{\bar{\ell}},\mathrm{b}_{\bar{m}}}[k]\right)^{*}}_{I_{2}}\nonumber\\
&+\underbrace{\sum_{k=0}^{N-1}\sum_{\imath'=N_{\mathrm{RF}}/2}^{N_{\mathrm{RF}}-1}\bm{w}^{*}_{m_{\mathrm{t}},r}\left(\sum_{r'=1}^{N_{\mathrm{r}}}\rho_{\tau_{r'}}[k]h_{r'}^{\mathrm{vh}}\bm{a}_{\mathrm{r}}(\psi_{r'})\bm{a}_{\mathrm{t}}^{*}(\theta_{r'},\phi_{r'})\right)\bm{a}_{\mathrm{t}}\left(\mu_{\mathrm{el}}',\mu_{\mathrm{az}}^{n_{\imath'}}\right)x^{\mathrm{a}_{\imath'},\mathrm{b}_{\imath'}}[k]\left(x^{\mathrm{a}_{\bar{\ell}},\mathrm{b}_{\bar{m}}}[k]\right)^{*}}_{I_{3}},\nonumber
\end{align}
where $u,\bar{u},\breve{u}\in\left\{0,\cdots,N_{\mathrm{RF}}/2-1\right\}$. In (\ref{i0i1i2}), $I_{0}$ denotes the correlation value if the azimuth transmit beamforming vector corresponds to the same auxiliary beam pair and paired-beam IDs as the reference ones, i.e., $\mathrm{a}_{u}=\mathrm{a}_{\bar{\ell}}$ and $\mathrm{b}_{u}=\mathrm{b}_{\bar{m}}$. The term $I_{1}$ represents the interference caused by the correlation when the azimuth transmit beamforming vector having the same auxiliary beam pair ID but different paired-beam ID with respect to the reference ones, i.e., $\mathrm{a}_{\bar{u}}=\mathrm{a}_{\bar{\ell}}$ and $\mathrm{b}_{\bar{u}}\neq\mathrm{b}_{\bar{m}}$. The interference term $I_{2}$ is obtained via the correlation when $\mathrm{a}_{\breve{u}}\neq\mathrm{a}_{\bar{\ell}}$ for either $\mathrm{b}_{\breve{u}}=\mathrm{b}_{\bar{m}}$ or $\mathrm{b}_{\breve{u}}\neq\mathrm{b}_{\bar{m}}$. The interference term $I_{3}$ is resulted from the design principle that beams formed from different polarization domains do not share the same auxiliary beam pair ID, i.e., $\mathrm{a}_{\imath'}\neq\mathrm{a}_{u} (\mathrm{a}_{\bar{\ell}})$ for either $\mathrm{b}_{\imath'}=\mathrm{b}_{u} (\mathrm{b}_{\bar{m}})$ or $\mathrm{b}_{\imath'}\neq\mathrm{b}_{u} (\mathrm{b}_{\bar{m}})$.

To compare the amplitudes of $I_{0}$, $I_{1}$, $I_{2}$ and $I_{3}$, their upper bounds are first derived in the absence of frequency selectivity, i.e., $\rho_{\tau_{r}}\triangleq\rho_{\tau_{r}}[0]=\cdots=\rho_{\tau_{r}}[N-1]$ for $r=1,\cdots,N_{\mathrm{r}}$. It can therefore be observed from (\ref{i0i1i2}) that the amplitude of $I_{0}$ is upper bounded by $N\sqrt{\frac{1}{1+\chi}}\big|\sum_{r'=1}^{N_{\mathrm{r}}}\rho_{\tau_{r'}}\\h_{r'}^{\mathrm{vv}}\big|$. The amplitude of $I_{1}$ can be upper bounded as
\begin{eqnarray}\label{up0}
\left|I_{1}\right|&\leq&\sqrt{\frac{1}{1+\chi}}\left|\sum_{r'=1}^{N_{\mathrm{r}}}\rho_{\tau_{r'}}h_{r'}^{\mathrm{vv}}\right|\left|\sum_{k=0}^{N-1}x^{\mathrm{a}_{\bar{u}},\mathrm{b}_{\bar{u}}}[k]\left(x^{\mathrm{a}_{\bar{\ell}},\mathrm{b}_{\bar{m}}}[k]\right)^{*}\right|\\
&=&\sqrt{\frac{1}{1+\chi}}\left|\sum_{r'=1}^{N_{\mathrm{r}}}\rho_{\tau_{r'}}h_{r'}^{\mathrm{vv}}\right|\left|x^{\mathrm{a}_{\bar{u}},\mathrm{b}_{\bar{u}}}[0]\left(x^{\mathrm{a}_{\bar{\ell}},\mathrm{b}_{\bar{m}}}[0]\right)^{*}\sum_{k=0}^{N-1}e^{\frac{j2\pi k\left(r^{\left(\mathrm{a}_{\bar{\ell}}\right)}p\left(\mathrm{b}_{\bar{u}}-\mathrm{b}_{\bar{m}}\right)\right)}{N}}\right|\\
&=&\Bigg\{\begin{array}{l}
            N\sqrt{\frac{1}{1+\chi}}\left|\sum_{r'=1}^{N_{\mathrm{r}}}\rho_{\tau_{r'}}h_{r'}^{\mathrm{vv}}\right|\left|x^{\mathrm{a}_{\bar{u}},\mathrm{b}_{\bar{u}}}[0]\left(x^{\mathrm{a}_{\bar{\ell}},\mathrm{b}_{\bar{m}}}[0]\right)^{*}\right|,\textrm{if}\hspace{1.5mm}r^{\left(\mathrm{a}_{\bar{\ell}}\right)}p\left(\mathrm{b}_{\bar{u}}-\mathrm{b}_{\bar{m}}\right)=\varrho N \\
            0,\hspace{2mm}\textrm{otherwise},
          \end{array}
\end{eqnarray}
where $\varrho$ is an arbitrary integer and $p$ is an arbitrary prime number. To achieve zero correlation interference, we can adjust $p$ such that $p\neq\frac{\varrho N}{r^{\left(\mathrm{a}_{\bar{\ell}}\right)}\left(\mathrm{b}_{\bar{u}}-\mathrm{b}_{\bar{m}}\right)}$ is achieved under the constraints of $p\leq p_{\mathrm{max}}$ and $\varrho\leq\varrho_{\mathrm{max}}$. Here, $p_{\mathrm{max}}$ denotes the maximum possible circular shift spacing in the frequency domain, i.e., $p_{\mathrm{max}}=\lfloor N/N_{\mathrm{p}}\rfloor$, where $N_{\mathrm{p}}$ represents the total number of possible paired-beam IDs. Further, denote by $\triangle\mathrm{b}_{\max}$ the maximum possible paired-beam ID difference, we have $\varrho_{\max}=\lceil r_{\mathrm{max}}p_{\mathrm{max}}\triangle\mathrm{b}_{\max}/N\rceil$ \cite{popov}, where $r_{\mathrm{max}}=\max\left\{r^{\left(\mathrm{a}_{\breve{\ell}}\right)},\breve{\ell}=0,\cdots,N_{\mathrm{a}}-1\right\}$. For our proposed multi-layer pilot structure, as $b_{m}\in\left\{0,1\right\}$ for $m\in\left\{0,\cdots,N_{\mathrm{az}}-1\right\}$, we therefore have $N_{\mathrm{p}}=2$ and $\triangle\mathrm{b}_{\max}=1$ resulting in $p_{\mathrm{max}}=N/2$ and $\varrho_{\mathrm{max}}=r_{\mathrm{max}}/2$.

Using the property of a Gauss sum \cite{bewg}, the amplitude of the correlation interference term $I_{2}$ is upper bounded as \cite{khcc}
\begin{eqnarray}\label{compee}
\left|I_{2}\right|&\leq&\sqrt{\frac{1}{1+\chi}}\left|\sum_{r'=1}^{N_{\mathrm{r}}}\rho_{\tau_{r'}}h_{r'}^{\mathrm{vv}}\right|\left|\sum_{\breve{u}\neq\bar{\ell}}\sum_{k=0}^{N-1}x^{\mathrm{a}_{\breve{u}},\mathrm{b}_{\breve{u}}}[k]\left(x^{\mathrm{a}_{\bar{\ell}},\mathrm{b}_{\bar{m}}}[k]\right)^{*}\right|.
\end{eqnarray}
The amplitude of $I_{2}$ can be further upper bounded by $\sqrt{\frac{1}{1+\chi}}\left|\sum_{r'=1}^{N_{\mathrm{r}}}\rho_{\tau_{r'}}h_{r'}^{\mathrm{vv}}\right|N_{\mathrm{e}}\sqrt{N}$ for either $\mathrm{b}_{\breve{u}}=\mathrm{b}_{\bar{m}}$ or $\mathrm{b}_{\breve{u}}\neq\mathrm{b}_{\bar{m}}$, where $N_{\mathrm{e}}$ denotes the total number of azimuth transmit steering vectors in $\bm{F}_{n_{\mathrm{t}}}$ formed from the vertically polarized antenna elements that have different auxiliary beam pair IDs from the reference auxiliary beam pair ID $\bar{\ell}$. That is, denote by $A=\left\{0,\cdots,N_{\mathrm{RF}}/2-1\right\}\setminus\bar{\ell}$, we therefore have $N_{\mathrm{e}}=\mathrm{card}\left(A\right)$, where $\mathrm{card}(\cdot)$ represents the cardinality of the set. Similarly, the amplitude of $I_{3}$ is upper bounded as
\begin{equation}\label{up3}
\left|I_3\right|\leq\sqrt{\frac{1}{1+\chi}}\left|\sum_{r'=1}^{N_{\mathrm{r}}}\rho_{\tau_{r'}}h_{r'}^{\mathrm{vh}}\right|\frac{N_{\mathrm{RF}}}{2}\sqrt{N},
\end{equation}
for either $\mathrm{b}_{\imath'}=\mathrm{b}_{\bar{m}}$ or $\mathrm{b}_{\imath'}\neq\mathrm{b}_{\bar{m}}$, where $\frac{N_{\mathrm{RF}}}{2}$ implies the total number of azimuth transmit steering vectors probed from the horizontally polarized antenna elements in $\bm{F}_{n_{\mathrm{t}}}$ (see (\ref{specase})). For large bandwidth at mmWave frequencies with large $N$ and limited number of RF chains, $\left|I_{0}\right|\gg\left|I_{2}\right|$ and $\left|I_{0}\right|\gg\left|I_{3}\right|$. We therefore have $\left|y^{u}_{m_{\mathrm{t}},r,n_{\mathrm{t}}}\right|\approx\left|I_{0}\right|$.

Denote by $\Omega_{r,u}=\sum_{k=0}^{N-1}\rho_{\tau_{r}}h_{r}^{\mathrm{vv}}[k]x^{\mathrm{a}_{u},\mathrm{b}_{u}}[k]\left(x^{\mathrm{a}_{u},\mathrm{b}_{u}}[k]\right)^{*}$, and
\begin{equation}\label{pathch}
\bm{\Lambda}_{r,u}=\sum_{k=0}^{N-1}\left(\sum_{r'=1, r'\neq r}^{N_{\mathrm{r}}}\rho_{\tau_{r'}}[k]h_{r'}^{\mathrm{vv}}\bm{a}_{\mathrm{r}}\left(\psi_{r'}\right)\bm{a}^{*}_{\mathrm{t}}\left(\theta_{r'},\phi_{r'}\right)\right)\left(x^{\mathrm{a}_{u},\mathrm{b}_{u}}[k]\right)^{*},
\end{equation}
where $u\in\left\{0,\cdots,N_{\mathrm{RF}}/2-1\right\}$. Further, we assume that the azimuth transmit spatial frequency $\mu_{\mathrm{y},r}$ for path-$r$ ($\mu_{\mathrm{y},r}=\frac{2\pi}{\lambda}d_{\mathrm{tx}}\sin(\theta_{r})\cos(\phi_{r})$) is within the range of $\left(\mu_{\mathrm{az}}^{n_{u}},\mu_{\mathrm{az}}^{n_{v}}\right)$ where $v\in\left\{0,\cdots,N_{\mathrm{RF}}/2-1\right\}$. $\bm{a}_{\mathrm{t}}\left(\mu_{\mathrm{el}}',\mu_{\mathrm{az}}^{n_{u}}\right)$ and $\bm{a}_{\mathrm{t}}\left(\mu_{\mathrm{el}}',\mu_{\mathrm{az}}^{n_{v}}\right)$ form an azimuth transmit auxiliary beam pair for a given elevation transmit direction $\mu_{\mathrm{el}}'$, and they come from the $n_{\mathrm{t}}$-th and ${s}_{\mathrm{t}}$-th azimuth probings by the BS. With respect to $\bm{a}_{\mathrm{t}}\left(\mu_{\mathrm{el}}',\mu_{\mathrm{az}}^{n_{u}}\right)$, the corresponding received signal strength can therefore be approximated as
\begin{align}
\left|y^{u}_{m_{\mathrm{t}},r,n_{\mathrm{t}}}\right|^2 \approx& \hspace{1mm}\label{last3} \left|\Omega_{r,u}\right|^2\left|\bm{w}^{*}_{m_{\mathrm{t}},r}\bm{a}_{\mathrm{r}}\left(\psi_{r}\right)\right|\left|\bm{a}_{\mathrm{t}}^{*}\left(\theta_{r},\phi_{r}\right)\bm{a}_{\mathrm{t}}\left(\mu_{\mathrm{el}}',\mu_{\mathrm{az}}^{n_{u}}\right)\right|\\
+&\hspace{1mm}\Omega_{r,u}^{*}\bm{w}^{*}_{m_{\mathrm{t}},r}\bm{\Lambda}^{*}_{r,u}\bm{a}_{\mathrm{t}}\left(\mu_{\mathrm{el}}',\mu_{\mathrm{az}}^{n_{u}}\right)\bm{a}_{\mathrm{t}}^{*}\left(\mu_{\mathrm{el}}',\mu_{\mathrm{az}}^{n_{u}}\right)\bm{a}_{\mathrm{t}}\left(\theta_{r},\phi_{r}\right)\bm{a}^{*}_{\mathrm{r}}\left(\psi_{r}\right)\bm{w}_{m_{\mathrm{t}},r}\nonumber\\
+&\hspace{1mm}\Omega_{r,u}\bm{w}^{*}_{m_{\mathrm{t}},r}\bm{a}_{\mathrm{r}}\left(\psi_{r}\right)\bm{a}_{\mathrm{t}}^{*}\left(\theta_{r},\phi_{r}\right)\bm{a}_{\mathrm{t}}\left(\mu_{\mathrm{el}}',\mu_{\mathrm{az}}^{n_{u}}\right)\bm{a}^{*}_{\mathrm{t}}\left(\mu_{\mathrm{el}}',\mu_{\mathrm{az}}^{n_{u}}\right)\bm{\Lambda}^{*}_{r,u}\bm{w}_{m_{\mathrm{t}},r}\nonumber\\
+&\hspace{1mm}\bm{w}^{*}_{m_{\mathrm{t}},r}\bm{\Lambda}_{r,u}\bm{a}_{\mathrm{t}}\left(\mu_{\mathrm{el}}',\mu_{\mathrm{az}}^{n_{u}}\right)\bm{a}^{*}_{\mathrm{t}}\left(\mu_{\mathrm{el}}',\mu_{\mathrm{az}}^{n_{u}}\right)\bm{\Lambda}^{*}_{r,u}\bm{w}_{m_{\mathrm{t}},r}\nonumber\\
\overset{\textrm{a.s.}}{\underset{N_{\mathrm{tot}}M_{\mathrm{tot}}\rightarrow\infty}{\xrightarrow{\hspace*{1.5cm}}}}&\hspace{1mm}\left|\Omega_{r,u}\right|^2\left|\bm{w}^{*}_{m_{\mathrm{t}},r}\bm{a}_{\mathrm{r}}\left(\psi_{r}\right)\right|\left|\bm{a}_{\mathrm{t}}^{*}\left(\theta_{r},\phi_{r}\right)\bm{a}_{\mathrm{t}}\left(\mu_{\mathrm{el}}',\mu_{\mathrm{az}}^{n_{u}}\right)\right|.\label{asymttt}
\end{align}
Note that (\ref{asymttt}) is obtained by exploiting the sparse nature of the mmWave channels such that if $N_{\mathrm{tot}}M_{\mathrm{tot}}\rightarrow\infty$, the last three terms in (\ref{last3}) converge to zero since $\bm{w}^{*}_{m_{\mathrm{t}},r}\bm{\Lambda}^{*}_{r,u}\bm{a}_{\mathrm{t}}\left(\mu_{\mathrm{el}}',\mu_{\mathrm{az}}^{n_{u}}\right)$ converges to zero \cite{omar2}. This is because for large numbers of transmit and receive antennas with angular sparsity, the projection of path-$r$'s beamforming and combining vectors $\bm{a}_{\mathrm{t}}\left(\mu_{\mathrm{el}}',\mu_{\mathrm{az}}^{n_{u}}\right)$ and $\bm{w}_{m_{\mathrm{t}},r}$ on path-$r'$'s ($r'\neq r$) channel $\bm{\Lambda}$ defined in (\ref{pathch}) becomes arbitrarily small. Similarly, for $\bm{a}_{\mathrm{t}}\left(\mu_{\mathrm{el}}',\mu_{\mathrm{az}}^{n_{v}}\right)$,
\begin{equation}\label{asymttx}
\left|y^{v}_{m_{\mathrm{t}},r,s_{\mathrm{t}}}\right|^2\overset{\textrm{a.s.}}{\underset{N_{\mathrm{tot}}M_{\mathrm{tot}}\rightarrow\infty}{\xrightarrow{\hspace*{1.5cm}}}}\hspace{1mm}\left|\Omega_{r,v}\right|^2\left|\bm{w}^{*}_{m_{\mathrm{t}},r}\bm{a}_{\mathrm{r}}\left(\psi_{r}\right)\right|\left|\bm{a}_{\mathrm{t}}^{*}\left(\theta_{r},\phi_{r}\right)\bm{a}_{\mathrm{t}}\left(\mu_{\mathrm{el}}',\mu_{\mathrm{az}}^{n_{v}}\right)\right|.
\end{equation}

Using the asymptotical results provided in (\ref{asymttt}) and (\ref{asymttx}), the corresponding ratio metric can be calculated as
\begin{eqnarray}
\zeta_{\mathrm{az},r}&=&\frac{\left|y^{u}_{m_{\mathrm{t}},r,n_{\mathrm{t}}}\right|^2-\left|y^{v}_{m_{\mathrm{t}},r,s_{\mathrm{t}}}\right|^2}{\left|y^{u}_{m_{\mathrm{t}},r,n_{\mathrm{t}}}\right|^2+\left|y^{v}_{m_{\mathrm{t}},r,s_{\mathrm{t}}}\right|^2}\label{denu}\\
&=&\frac{\left|\Omega_{r,u}\right|^2\left|\bm{a}_{\mathrm{t}}^{*}\left(\theta_{r},\phi_{r}\right)\bm{a}_{\mathrm{t}}\left(\mu_{\mathrm{el}}',\mu_{\mathrm{az}}^{n_{u}}\right)\right|-\left|\Omega_{r,v}\right|^2\left|\bm{a}_{\mathrm{t}}^{*}\left(\theta_{r},\phi_{r}\right)\bm{a}_{\mathrm{t}}\left(\mu_{\mathrm{el}}',\mu_{\mathrm{az}}^{n_{v}}\right)\right|}{\left|\Omega_{r,u}\right|^2\left|\bm{a}_{\mathrm{t}}^{*}\left(\theta_{r},\phi_{r}\right)\bm{a}_{\mathrm{t}}\left(\mu_{\mathrm{el}}',\mu_{\mathrm{az}}^{n_{u}}\right)\right|+\left|\Omega_{r,v}\right|^2\left|\bm{a}_{\mathrm{t}}^{*}\left(\theta_{r},\phi_{r}\right)\bm{a}_{\mathrm{t}}\left(\mu_{\mathrm{el}}',\mu_{\mathrm{az}}^{n_{v}}\right)\right|} \label{canr}\\
&=&-\frac{\sin\left(\frac{\mu_{\mathrm{y},r}-\mu_{\mathrm{az}}}{2}\right)\sin\left(\delta_{\mathrm{y}}\right)}{1-\cos\left(\mu_{\mathrm{y},r}-\mu_{\mathrm{az}}\right)\cos\left(\delta_{\mathrm{y}}\right)}\label{clra},
\end{eqnarray}
where $\delta_{\mathrm{y}}=\left|\mu_{\mathrm{az}}^{n_{v}}-\mu_{\mathrm{az}}^{n_{u}}\right|/2$, and $\mu_{\mathrm{az}}=\mu_{\mathrm{az}}^{n_{v}}-\delta_{\mathrm{y}}$ or $\mu_{\mathrm{az}}=\mu_{\mathrm{az}}^{n_{u}}+\delta_{\mathrm{y}}$.
The azimuth transmit spatial frequency for path-$r$ can therefore be estimated as
\begin{eqnarray}\label{aoanglepathr}
\hat{\mu}_{\mathrm{y},r}=\mu_{\mathrm{az}}-\arcsin\Bigg(\frac{\zeta_{\mathrm{az},r}\sin(\delta_{\mathrm{y}})-\zeta_{\mathrm{az},r}\sqrt{1-\zeta_{\mathrm{az},r}^{2}}\sin(\delta_{\mathrm{y}})\cos(\delta_{\mathrm{y}})}{\sin^{2}(\delta_{\mathrm{y}})+\zeta_{\mathrm{az},r}^{2}\cos^{2}(\delta_{\mathrm{y}})}\Bigg).
\end{eqnarray}

The custom designed multi-layer pilot structure is able to reduce pilot resource overhead. Here, the pilot resource is defined as the required least number of ZC sequences that have different root indices. An alternative to the proposed multi-layer pilot design is the conventional single-layer pilot structure such that each beam is simply associated with a distinct ZC sequence with a unique root index. For the proposed multi-layer structure, however, only the auxiliary beam pair is associated with a distinct ZC sequence with two fixed shift spacing values corresponding to the two beams in the same pair. For instance, assume that $N_{\mathrm{RF}}$ transmit beams are simultaneously probed. For the single-layer pilot design, at least $N_{\mathrm{RF}}$ ZC sequences with different root indices are needed. Regarding the multi-layer structure, this number becomes $N_{\mathrm{RF}}/2$, which reduces the pilot resource overhead by $50\%$. Reducing the pilot resource overhead is beneficial such that appropriate ZC sequences (with different root indices) with good correlation properties such as those employed in LTE PSS (ZC sequences with root indices 25, 29 and 34) can be flexibly selected from the sequence pool to improve beam detection performance in wideband channels.
\subsubsection{Practical implementation for the proposed approach}
As can be seen from the asymptotical results derived in (\ref{asymttt}) and (\ref{asymttx}), the ratio metric for estimating path-$r$'s azimuth transmit spatial frequency does not depend on the receive combining vector and elevation steering direction. In communications systems, however, the angle estimation performance is subject to the noise power, which makes the selection of appropriate auxiliary beam pairs important for practical implementation. In the following, the procedures of selecting proper transmit auxiliary beam pairs for multi-path's AoD estimation are illustrated.

A receive probing matrix is first chosen such that the sum of the received signal strengths with respect to all of its contained receive combining vectors is maximized among all receive probing matrices. With respect to the selected receive probing matrix, a predefined number of transmit beamforming vectors that yield the largest received signal strengths are selected. The number of selected transmit beamforming vectors ought to be greater than or equal to the number of channel paths, though it depends on practical implementation. According to \cite[Lemma~2]{dztrans}, the corresponding elevation/azimuth transmit auxiliary beam pairs that comprise of the selected transmit beamforming vectors cover the elevation/azimuth AoDs to be estimated with high probability. For a given selected transmit beamforming vector, the received signal strengths of its two adjacent beams in the elevation/azimuth angular domain are compared, and the adjacent beam with the highest received signal strength among the two is then selected and paired with the previously selected transmit beamforming vector as the elevation/azimuth transmit auxiliary beam pair of interest. The ratio metric with respect to each determined elevation/azimuth transmit auxiliary beam pair is then computed, which determines the corresponding elevation/azimuth transmit spatial frequencies using (\ref{aoanglepathr}). For the multi-RF case, the computational complexity of the entire process can then be computed as $N_{\mathrm{RF}}N_{\mathrm{TX}}M_{\mathrm{RF}}M_{\mathrm{RX}}$, where $N_{\mathrm{TX}}$ represents the total number of transmit probings in both azimuth and elevation domains, and $M_{\mathrm{RX}}$ denotes the total number of receive probings.

\section{Differential Feedback for Auxiliary Beam Pair Design}
In a closed-loop FDD system, the acquisition of the azimuth/elevation AoD at the BS requires information feedback from the UE via a feedback channel. In our prior work in \cite{dztrans}, direct transmit spatial frequency and ratio metric quantization and feedback strategies were discussed. In this paper, a differential quantization and feedback option is custom designed for the proposed algorithm to reduce the feedback overhead in MIMO systems with two-dimensional phased array. For simplicity, the narrowband example employed in Section III-A is used throughout this section. In addition, we only consider transmit beamforming in the azimuth domain. For general wideband channels with multi-path angle components and two-dimensional phased array, both the estimated azimuth and elevation AoDs for each path are quantized and fed back.


In fact, the sign of the ratio metric implies the relative position of the AoD/AoA to the boresight of the corresponding auxiliary beam pair. A conceptual example showing the sign effect of the ratio metric is given in Fig.~4 using one azimuth transmit auxiliary beam pair. In Fig.~4(a), the azimuth transmit spatial frequency $\mu_{\mathrm{y}}$ is to the left of the boresight of the auxiliary beam pair $\mu_{\mathrm{az}}$ such that $\mu_{\mathrm{y}}\in\left(\mu_{\mathrm{az}}-\delta_{\mathrm{y}},\mu_{\mathrm{az}}\right)$. In fact, it can be seen from (\ref{adhoc}) that, the sign of the azimuth ratio metric purely depends on the sign of $\sin\left(\frac{\mu_{\mathrm{y}}-\mu_{\mathrm{az}}}{2}\right)$. Since in this example $\mu_{\mathrm{y}}\in\left(\mu_{\mathrm{az}}-\delta_{\mathrm{y}},\mu_{\mathrm{az}}\right)$ and $0\leq\delta_{\mathrm{y}}<\pi/2$, $\zeta_{\mathrm{az}}$ is therefore positive such that $\mathrm{sign}\left(\zeta_{\mathrm{az}}\right)=1$. Similarly, in Fig.~4(b), as $\mu_{\mathrm{y}}\in\left(\mu_{\mathrm{az}},\mu_{\mathrm{az}}+\delta_{\mathrm{y}}\right)$, i.e., $\mu_{\mathrm{y}}$ is to the right of $\mu_{\mathrm{az}}$, $\mathrm{sign}\left(\zeta_{\mathrm{az}}\right)=-1$.
\begin{figure}
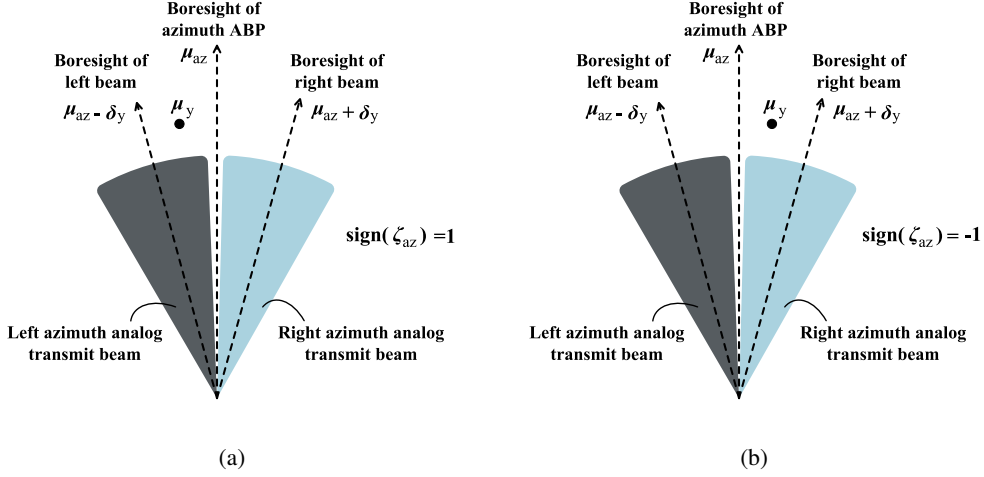

\begin{center}
\subfigure[]{%
\includegraphics[width=2.45in]{Fig_sign_left.pdf}
\label{fig:subfigure1}}
\quad
\subfigure[]{%
\includegraphics[width=2.45in]{Fig_sign_right.pdf}
\label{fig:subfigure2}}
\caption{A conceptual example of the relationship between the relative position of the azimuth transmit spatial frequency to the boresight of the corresponding azimuth transmit auxiliary beam pair and the sign of the azimuth ratio metric: (a) $\mu_{\mathrm{y}}\in\left(\mu_{\mathrm{az}}-\delta_{\mathrm{y}},\mu_{\mathrm{az}}\right)$, $\mathrm{sign}\left(\zeta_{\mathrm{az}}\right)=1$. (b) $\mu_{\mathrm{y}}\in\left(\mu_{\mathrm{az}},\mu_{\mathrm{az}}+\delta_{\mathrm{y}}\right)$, $\mathrm{sign}\left(\zeta_{\mathrm{az}}\right)=-1$.}
\label{fig:figure}
\end{center}
\end{figure}

With the knowledge of $\mu_{\mathrm{az}}$ and $\delta_{\mathrm{y}}$, the UE can first obtain $\hat{\mu}_{\mathrm{y}}$ according to (\ref{elaoangle}). As $\mu_{\mathrm{az}}$ and $\delta_{\mathrm{y}}$ are predetermined semi-static parameters, they can be periodically broadcasted from the BS to the UE. The difference between the estimated azimuth transmit spatial frequency and the boresight of the corresponding azimuth transmit auxiliary beam pair can then be determined as $\triangle\hat{\mu}_{\mathrm{y}}=|\hat{\mu}_{\mathrm{y}}-\mu_{\mathrm{az}}|$. Note that now $\triangle\hat{\mu}_{\mathrm{y}}\in\left[-\delta_{\mathrm{y}},\delta_{\mathrm{y}}\right]$. A relatively small codebook with codewords uniformly distributed within the interval of $\left[-\delta_{\mathrm{y}},\delta_{\mathrm{y}}\right]$ can therefore be used to quantize $\triangle\hat{\mu}_{\mathrm{y}}$. Along with the feedback of $\mathrm{sign}\left(\zeta_{\mathrm{az}}\right)$ ($1$ bit indicating either "$1$" or "$-1$"), the BS can retrieve the azimuth transmit spatial frequency as  $\hat{\mu}_{\mathrm{y}}=\mu_{\mathrm{az}}+\mathrm{sign}\left(\zeta_{\mathrm{az}}\right)\triangle\hat{\mu}_{\mathrm{y}}$.
\section{Numerical Results}
In this section, the performance of the proposed auxiliary beam pair enabled two-dimensional angle estimation technique is evaluated. The BS and UE employ the UPA and ULA with inter-element spacing of $\lambda/2$ between co-polarized antennas. We set $\delta_{\mathrm{x}}=\frac{\pi}{2N_{\mathrm{x}}}$, $\delta_{\mathrm{y}}=\frac{\pi}{2N_{\mathrm{y}}}$ and $\delta_{\mathrm{r}}=\frac{\pi}{2M_{\mathrm{tot}}}$ to approximate the half-power beamwidth of the corresponding beamforming and combining vectors. The BS covers $120^{\circ}$ angular range $\left[-60^\circ,60^\circ\right]$ around azimuth boresight ($0^\circ$) and $90^{\circ}$ angular range $\left[-45^{\circ},45^\circ\right]$ around elevation boresight ($0^\circ$). The UE monitors $180^{\circ}$ angular region $\left[-90^\circ,90^\circ\right]$ around boresight ($0^\circ$). The differential transmit spatial frequency feedback strategy is employed. The codewords for quantizing the azimuth and elevation AoDs are uniformly distributed within $\left[-\delta_{\mathrm{y}},\delta_{\mathrm{y}}\right]$ and $\left[-\delta_{\mathrm{x}},\delta_{\mathrm{x}}\right]$, and $3$ bits feedback is implemented for each angular domain.
\subsection{Narrowband single-path angle estimation using single RF chain without cross-polarization}
The azimuth/elevation AoD and AoA are assumed to take continuous values, i.e., not quantized, and are uniformly distributed within the corresponding coverage ranges. The beam codebook size for the azimuth transmit domain, elevation transmit domain and receive domain is determined as $\lceil 120^\circ/2\delta_{\mathrm{y}}\rceil$, $\lceil90^\circ/2\delta_{\mathrm{x}}\rceil$ and $\lceil180^{\circ}/2\delta_{\mathrm{r}}\rceil$ to avoid coverage holes. For instance, for $M_{\mathrm{tot}}=4$, the codebook size for the UE is $\lceil180^\circ/2\delta_{\mathrm{r}}\rceil=\lceil180^\circ/45^\circ\rceil=4$. A Rician channel model is employed throughout this part without considering cross-polarization. Specifically, for a given Rician $\mathsf{K}$-factor value,
\begin{align}
\bm{H} = \sqrt{\frac{\mathsf{K}}{1+\mathsf{K}}}\underbrace{g_{r}\bm{a}_{\mathrm{r}}(\psi_{r})\bm{a}_{\mathrm{t}}^{*}(\theta_{r},\phi_r)}_{\bm{H}_{\mathrm{LOS}}}+\sqrt{\frac{\mathsf{1}}{1+\mathsf{K}}}\underbrace{\sum_{r'=1,r'\neq r}^{N_{\mathrm{p}}}g_{r'}\bm{a}_{\mathrm{r}}(\psi_{r'})\bm{a}_{\mathrm{t}}^{*}(\theta_{r'},\phi_{r'})}_{\bm{H}_{\mathrm{NLOS}}},
\end{align}
where $\bm{H}_{\mathrm{LOS}}$ and $\bm{H}_{\mathrm{NLOS}}$ represent line-of-sight (LOS) and non-LOS (NLOS) channel components. We assume the number of NLOS channel components is $5$. The objective is to estimate the dominant LOS path's AoD and AoA. From the channel measurements in \cite{rician}, we set $13.2$dB Rician $\mathsf{K}$-factor that characterizes the mmWave channel in an urban wireless channel topography.

\begin{figure}
\centering
\subfigure[]{%
\includegraphics[width=2.71in]{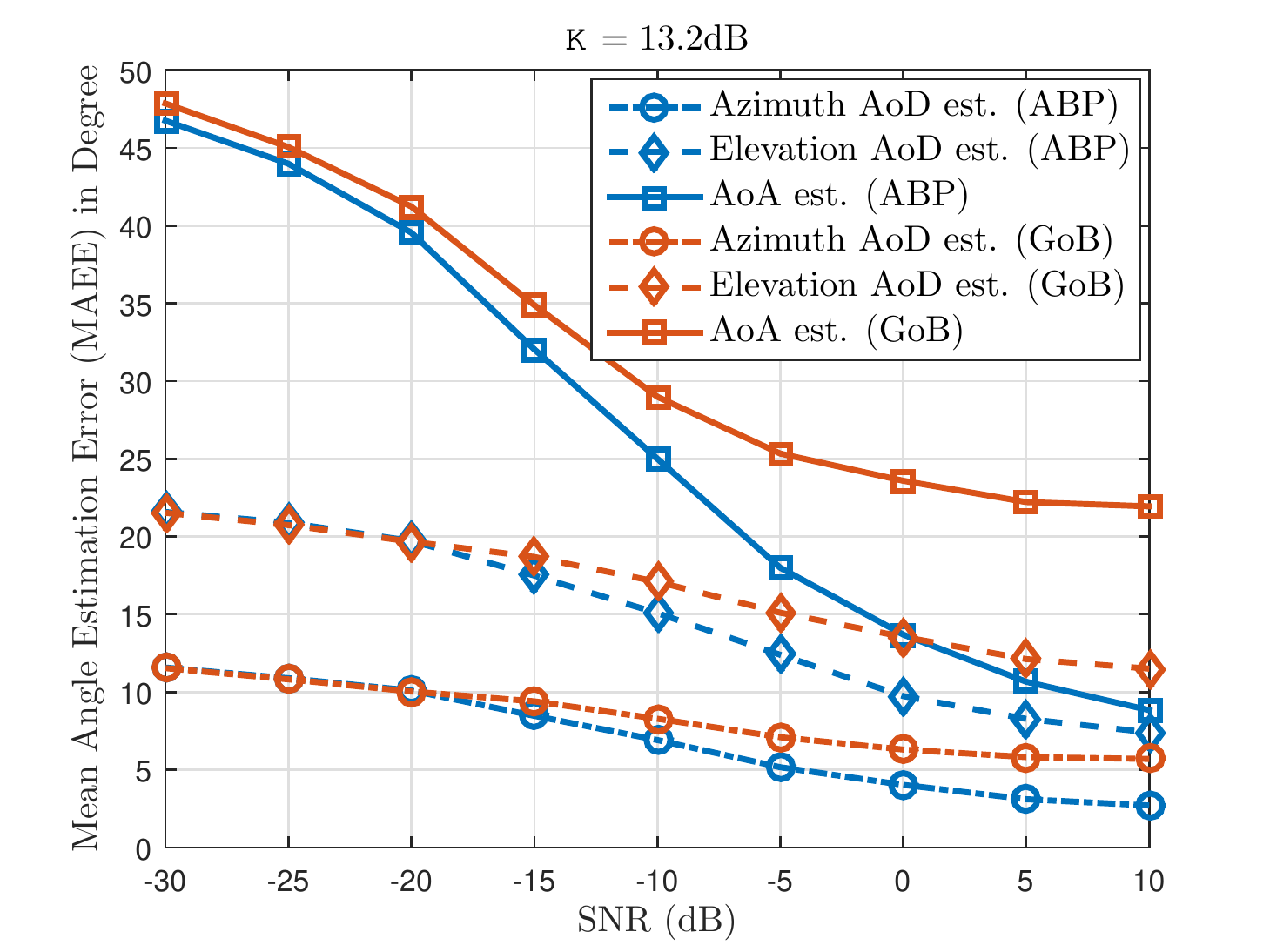}
\label{fig:subfigure3}}
\quad
\subfigure[]{%
\includegraphics[width=2.71in]{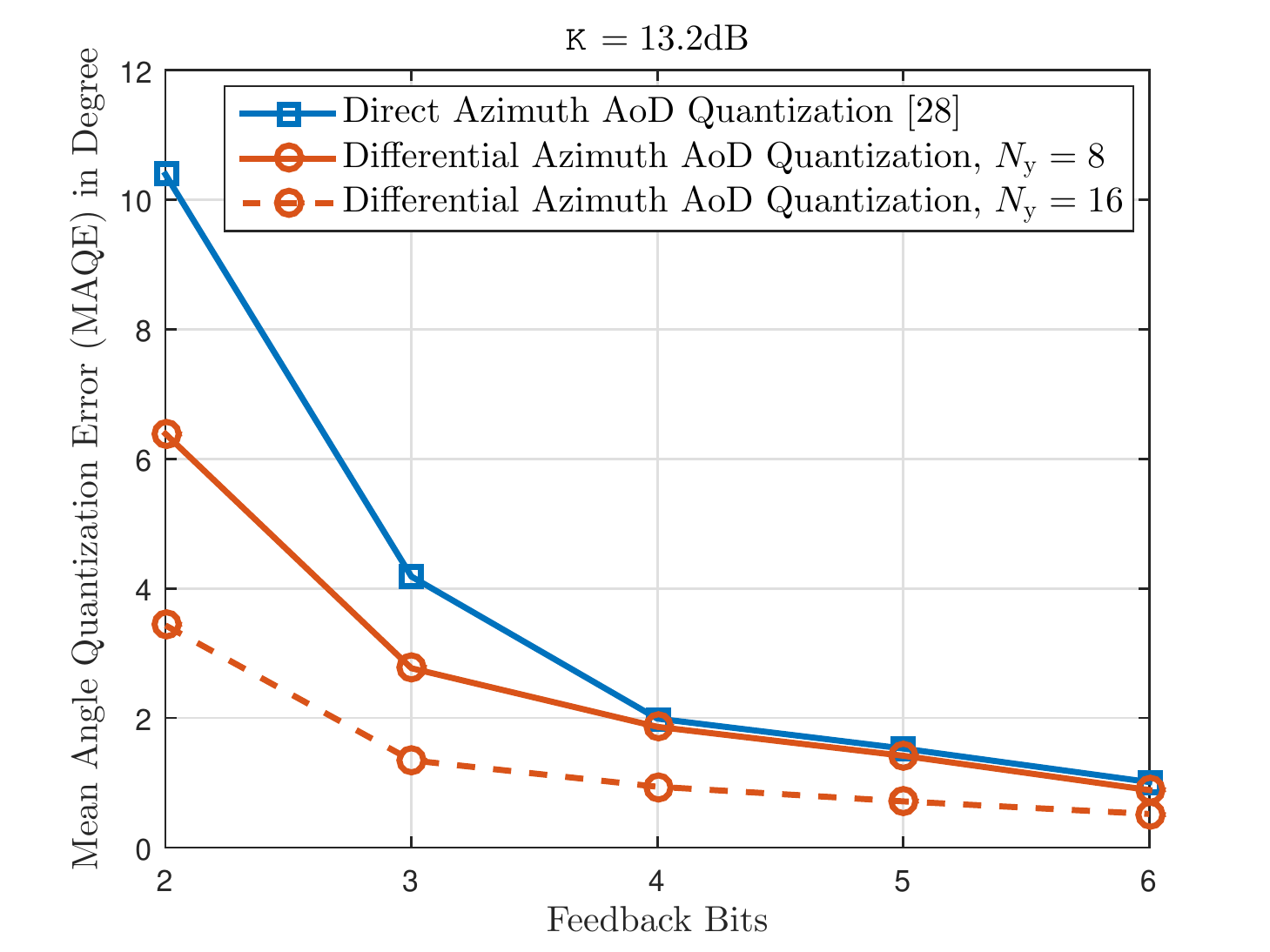}
\label{fig:subfigure3}}
\caption{(a) Mean angle estimation error (MAEE) performance of azimuth/elevation AoD and AoA estimation using the proposed auxiliary beam pair (ABP) and grid-of-beams (GoB) based methods. (b) Mean angle quantization error (MAQE) performance of quantizing the azimuth AoD using direct quantization in \cite{dztrans} and the newly proposed differential quantization.}
\label{fig:figure}
\end{figure}
The mean angle estimation error (MAEE) performance of the single-path's azimuth/elevation AoD and AoA acquisition is provided in Fig.~5(a) under various target SNR levels $\gamma$. The grid-of-beams based approach \cite{singh2} is also evaluated for comparison. Here, the MAEE is defined as $\mathbb{E}\left[\left|\upsilon_{\mathrm{true}}-\upsilon_{\mathrm{est}}\right|\right]$, where $\upsilon_{\mathrm{true}}$ represents the exact angle in degree, and $\upsilon_{\mathrm{est}}$ is its estimated counterpart in degree. Further, $N_{\mathrm{x}}=4, N_{\mathrm{y}}=8$, and $M_{\mathrm{tot}}=4$ are assumed. It is observed from Fig.~5(a) that promising MAEE performance of azimuth/elevation AoD and AoA estimation via the proposed auxiliary beam pair design can be achieved even at relatively low SNR regime. It can also be observed from Fig.~5(a) that the proposed technique outperforms the grid-of-beams based method for various target SNR levels.

In Fig.~5(b), the direct and differential AoD quantization and feedback strategies are compared in terms of the mean angle quantization error (MAQE). The MAQE is defined as $\mathbb{E}\left[\left|\upsilon_{\mathrm{est}}-\upsilon_{\mathrm{quan}}\right|\right]$, where $\upsilon_{\mathrm{quan}}$ represents the quantized version of the estimated value $\upsilon_{\mathrm{est}}$ in degree. Only the azimuth AoD quantization is examined in this example assuming $10$dB SNR. For the direct quantization, the codewords are uniformly distributed within the interval of $\left[-60^{\circ},60^{\circ}\right]$. Regarding the proposed differential approach, the codewords are uniformly distributed within $\left[-\delta_{\mathrm{y}},\delta_{\mathrm{y}}\right]$ where $\delta_{\mathrm{y}}=11.25^\circ$ and $\delta_{\mathrm{y}}=5.62^\circ$ for $N_{\mathrm{y}}=8$ and $N_{\mathrm{y}}=16$. Note that one extra bit indicating the sign is added to the number of feedback bits required by the differential approach. From Fig.~5(b), it is observed that using the same amount of feedback bits, the proposed differential approach exhibits better MAQE performance than the direct quantization. That is, for a given target quantization error, the differential strategy requires less amount of feedback overhead than the direct quantization approach.
\subsection{Wideband multi-path angle estimation using multiple RF chains with cross-polarization}
In this part of simulation, the statistical mmWave channel model developed in \cite{kssr} is implemented using the NYUSIM open source platform. The urban micro-cellular (UMi) scenario with NLOS components is considered for $28$GHz carrier frequency. Both $125$MHz and $250$MHz RF bandwidths are evaluated with $N=512$ and $N=1024$ subcarriers. The corresponding CP lengths are $D=64$ and $D=256$. The subcarrier spacing and symbol duration are set to $270$KHz and $3.7$us following the numerology provided in \cite{jerrypikhan}. The channel is modeled as a clustered channel where each cluster comprises several subpaths. Detailed channel modeling parameters including the distributions of clusters, subpaths in each cluster, azimuth/elevation AoD and AoA, the corresponding root-mean-square delay spreads and etc. are given in \cite[TABLE~III]{kssr}. The root indices for generating the first-layer auxiliary beam pair specific ZC sequences are chosen from $\left\{1,2,\cdots,N-1\right\}$, and they are pairwise primes with respect to $N-1$. The actual lengths of the ZC sequence are set to $511$ and $1023$ corresponding to the $125$MHz and $250$MHz bandwidths with the IFFT sizes of $512$ and $1024$, and the direct current (DC) subcarrier is set to zero. The frequency circular shift spacing $p$ is configured as $6$.

\begin{figure}
\centering
\subfigure[]{%
\includegraphics[width=2.71in]{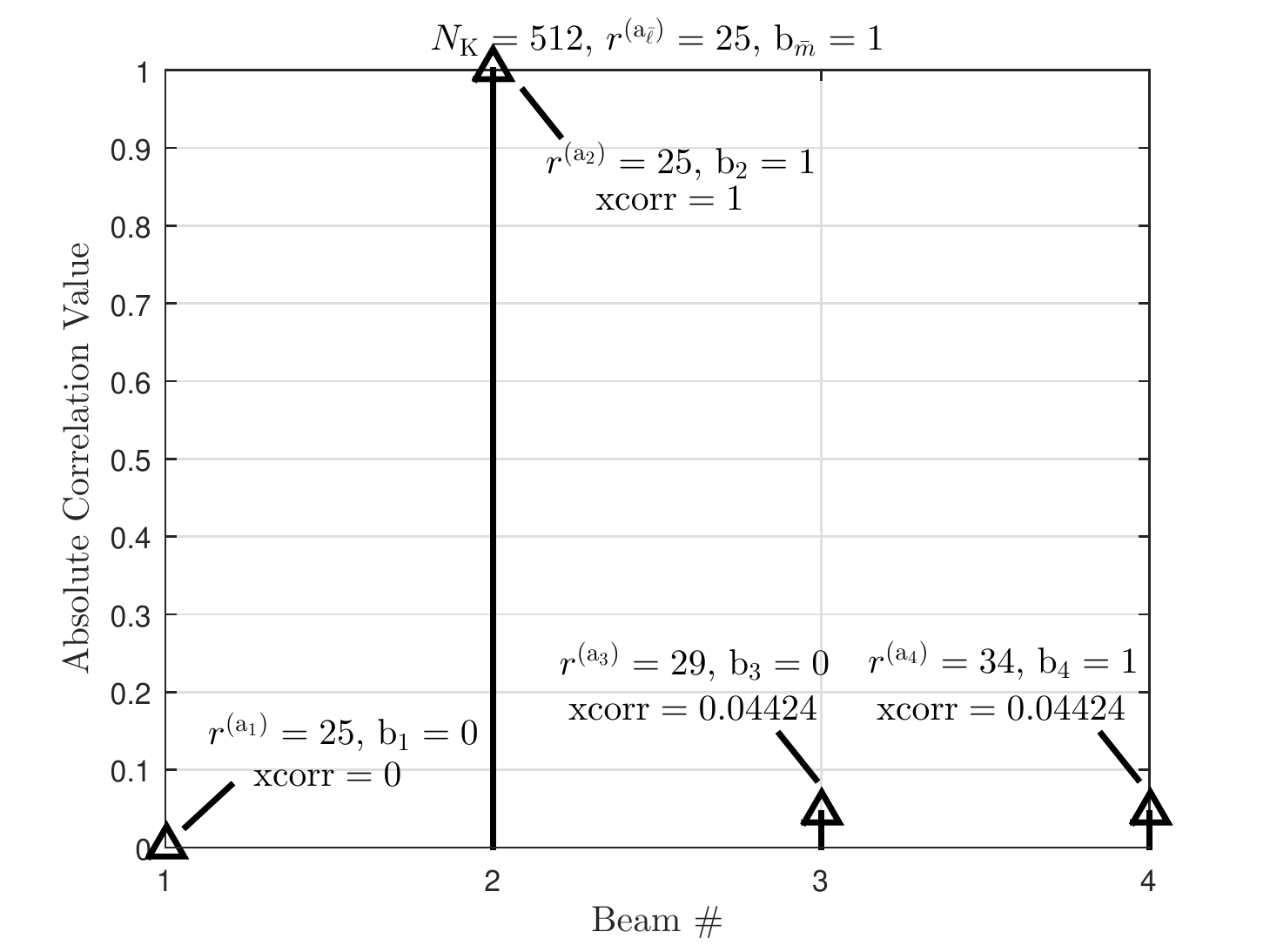}
\label{fig:subfigure3}}
\quad
\subfigure[]{%
\includegraphics[width=2.71in]{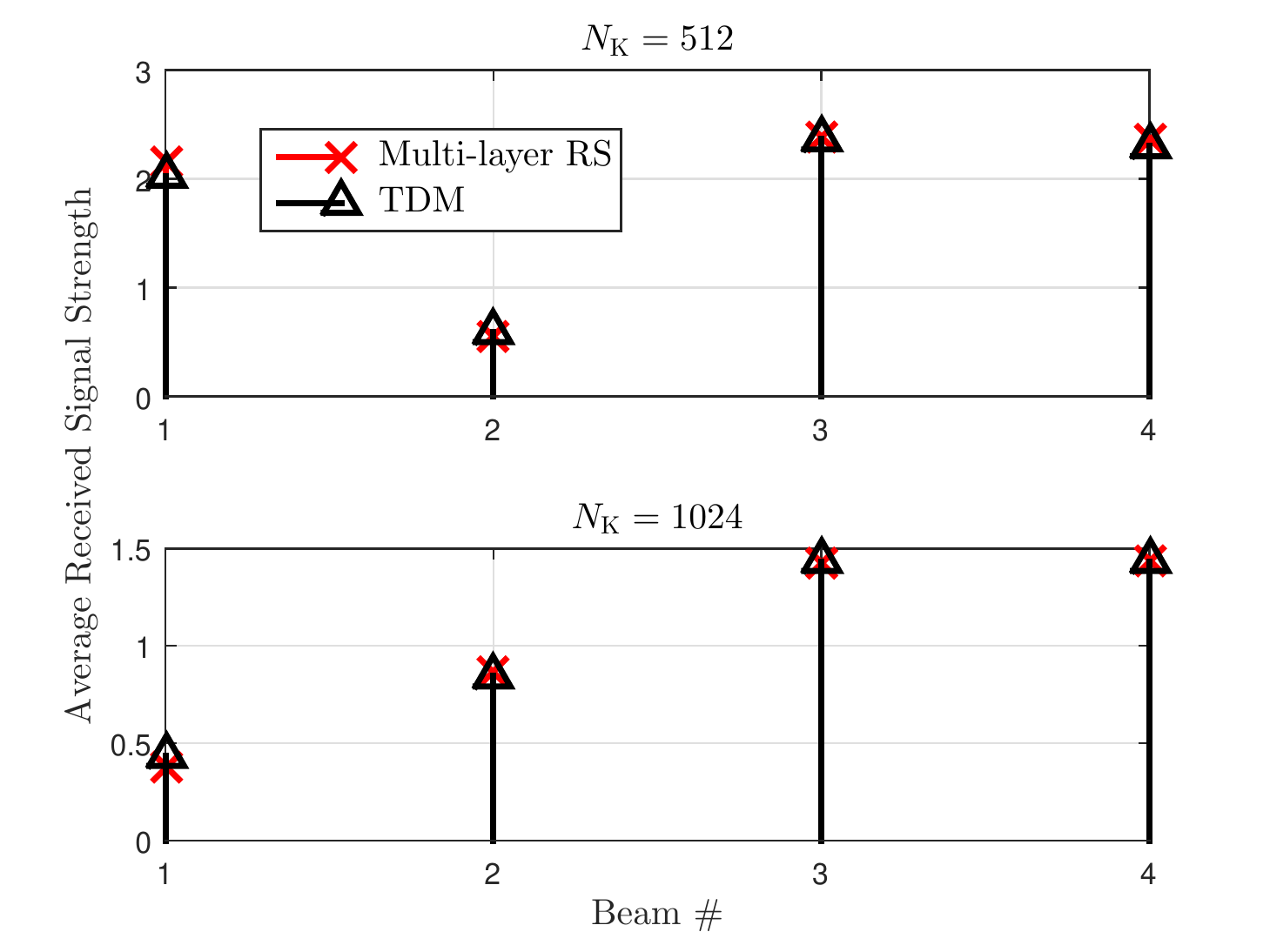}
\label{fig:subfigure3}}
\caption{(a) Absolute correlation value for multi-layer pilot design. (b) Average received signal strengths of simultaneously probed transmit beams using multi-layer pilot design. Cross-polarized antenna setup is evaluated with $N_{\mathrm{x}}=4$, $N_{\mathrm{y}}=8$, $N_{\mathrm{tot}}=64$, $M_{\mathrm{tot}}=8$, $\zeta=20^\circ$ and $\chi=0.2$.}
\label{fig:figure}
\end{figure}
We first examine the correlation property of the proposed multi-layer pilot design. In this example, four transmit beams are considered, and they are sequentially indexed from $1\sim4$. Their auxiliary beam pair IDs and paired-beam IDs are set to $r^{(\mathrm{a}_{1})}=r^{(\mathrm{a}_{2})}=25$, $r^{(\mathrm{a}_{3})}=29$, $r^{(\mathrm{a}_{4})}=34$, and $\mathrm{b}_{1}=0$, $\mathrm{b}_{2}=1$, $\mathrm{b}_{3}=0$, $\mathrm{b}_{4}=1$. The multi-layer pilots are correspondingly formed using $\left\{r^{(\mathrm{a}_{1})},\mathrm{b}_{1}\right\}$, $\left\{r^{(\mathrm{a}_{2})},\mathrm{b}_{2}\right\}$, $\left\{r^{(\mathrm{a}_{3})},\mathrm{b}_{3}\right\}$ and $\left\{r^{(\mathrm{a}_{4})},\mathrm{b}_{4}\right\}$ for beams $\#1$, $\#2$, $\#3$ and $\#4$. In Fig.~6(a), the absolute correlation values with respect to the four ZC sequences are plotted. The absolute correlation value is calculated as
\begin{equation}\label{xcorr}
\mathrm{xcorr} = \left|\frac{1}{N}\sum_{k=0}^{N-1}x^{\mathrm{a}_{\ell},\mathrm{b}_{m}}[k]\left(x^{\mathrm{a}_{\bar{\ell}},\mathrm{b}_{\bar{m}}}[k]\right)^{*}\right|,\hspace{2mm}\ell,m=1,2,3,4,
\end{equation}
where $r^{(\mathrm{a}_{\bar{\ell}})}=25$ and $\mathrm{b}_{\bar{m}}=1$ are the reference auxiliary beam pair ID and paired-beam ID. Note that only the zero-lag absolute correlation value is calculated assuming perfect time-frequency synchronization. This is different from the PSS-based OFDM symbol timing synchronization by window sliding, in which complete auto- and cross-correlation values are examined. It can be observed from Fig.~6(a) that as $r^{(\mathrm{a}_{2})}=r^{(\mathrm{a}_{\bar{\ell}})}$ and $\mathrm{b}_{2}=\mathrm{b}_{\bar{m}}$, the absolute correlation for beam $\#2$ exhibits the largest value, i.e., $1$. Correspondingly, the absolute correlations for beams $\#1$, $\#3$ and $\#4$ are $0$, $0.4424$ and $0.4424$. These observations are consistent with the analysis provided in (\ref{up0})-(\ref{up3}). In the example shown in Fig.~6(b), the four transmit beams are simultaneously probed using four transmit RF chains. The $125$MHz wideband channel with IFFT size of $512$ is evaluated. The first three transmit beams are chosen from the vertical transmit beam codebook while the last one is selected from the horizontal transmit beam codebook. Only one receive beam is considered, which is randomly selected from the vertical receive beam codebook. The received signal strength is plotted in Fig.~6(b) by averaging over all subcarriers. We also evaluate the case that all four beams are probed in a round-robin TDM fashion using one transmit RF chain without applying any pilot. The TDM case here can be treated as the ideal case such that the four beams can be perfectly identified because of the orthogonality in the time domain. In this example, by comparing with the TDM case, the capability of the proposed multi-layer pilot design in differentiating simultaneously probed beams in the code domain is examined. It can be seen from Fig.~6(b) that by decoupling each transmit beam with the corresponding reference ZC sequences, the average receive signal strengths in the proposed multi-layer pilot design are almost the same as the TDM case for all four beams.

\begin{figure}
\centering
\subfigure[]{%
\includegraphics[width=2.71in]{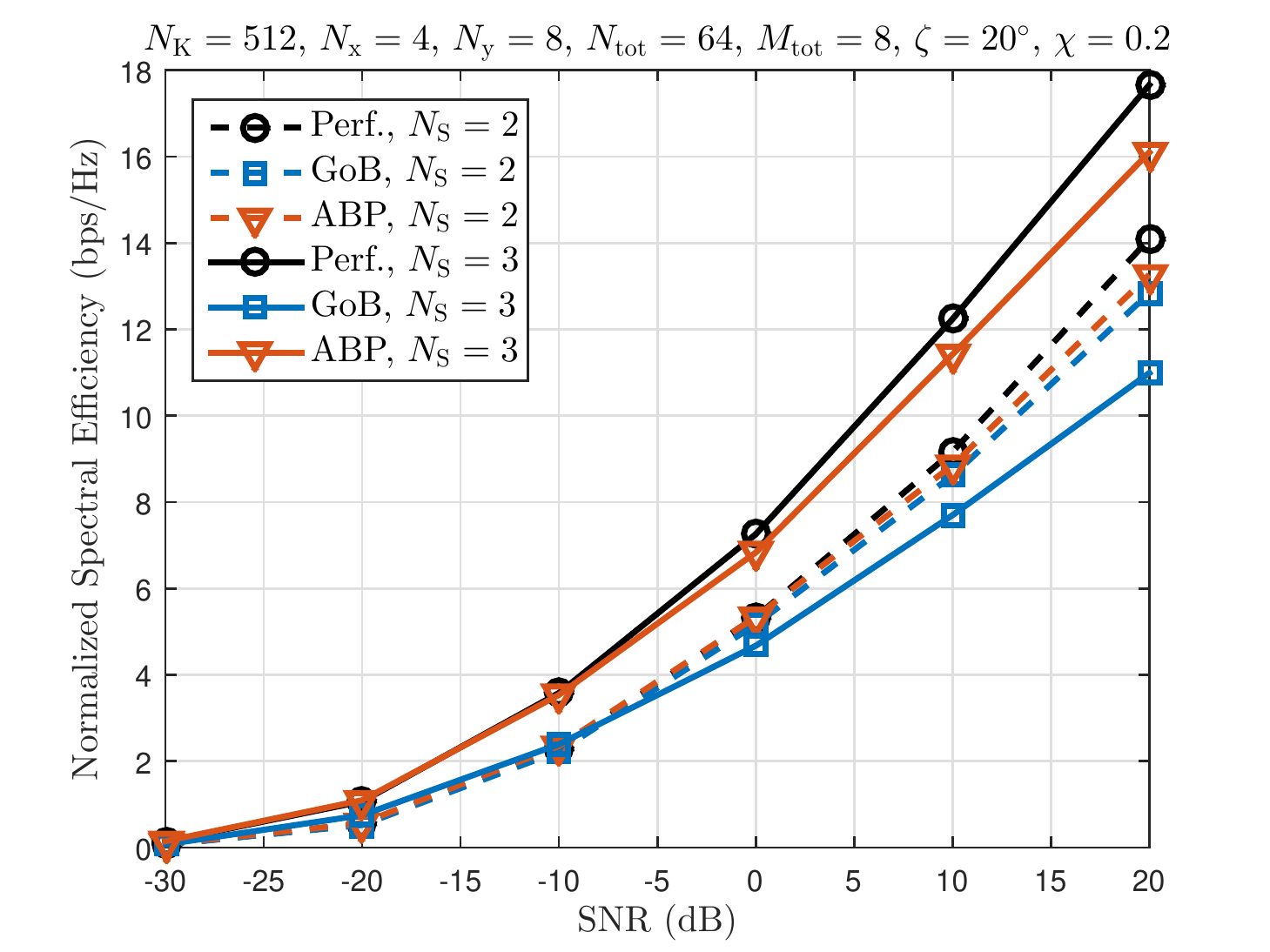}
\label{fig:subfigure3}}
\quad
\subfigure[]{%
\includegraphics[width=2.71in]{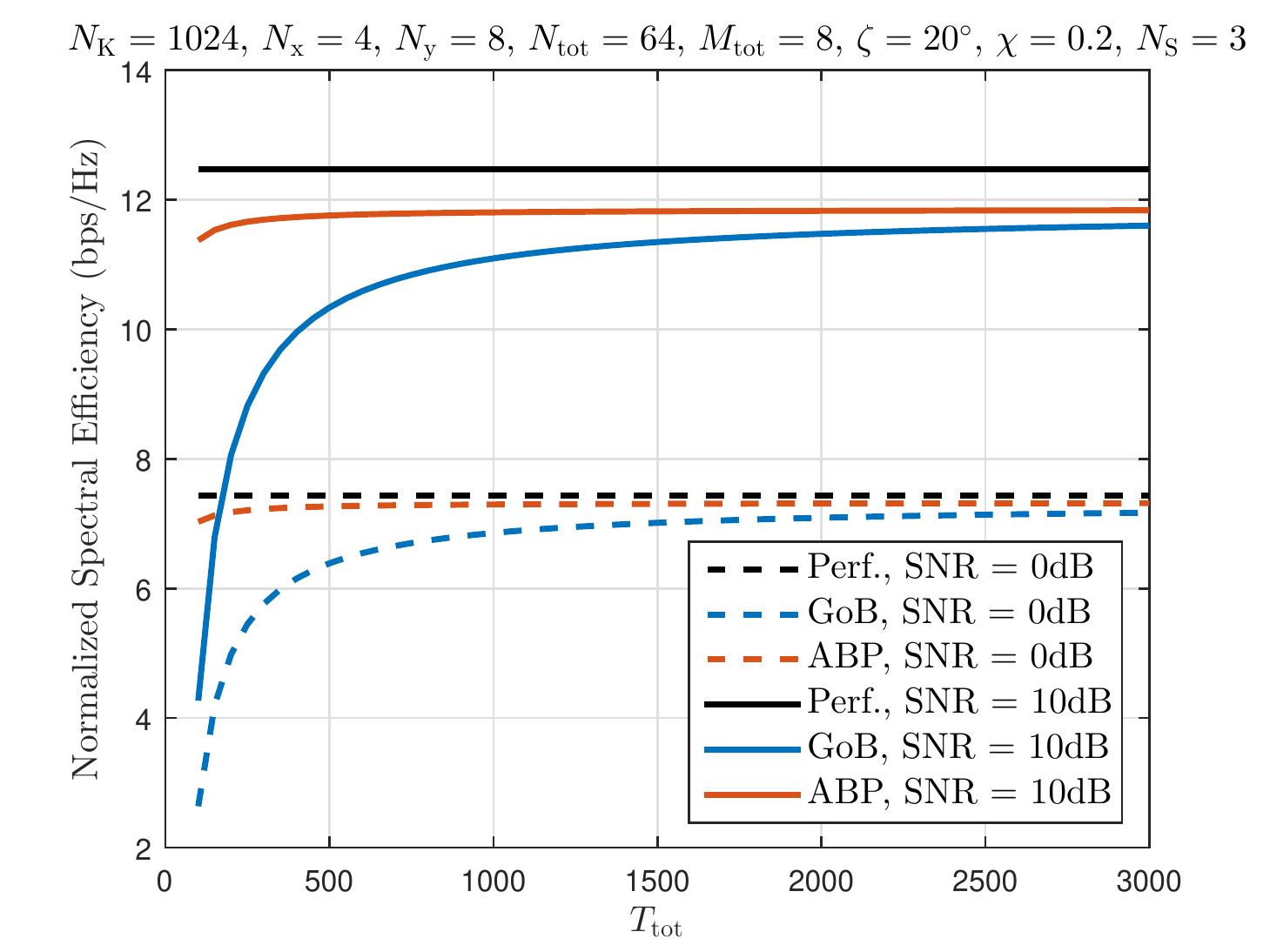}
\label{fig:subfigure3}}
\caption{(a) Normalized spectral efficiency of multi-layer transmission ($N_{\mathrm{S}}=2,3$) using perfect and estimated azimuth/elevation AoD and AoA via the proposed auxiliary beam pair and grid-of-beams based designs for $125$MHz bandwidth at various SNR levels. (b) Normalized spectral efficiency of multi-layer transmission ($N_{\mathrm{S}}=3$) using perfect and estimated azimuth/elevation AoD and AoA via the proposed auxiliary beam pair and grid-of-beams based designs for $250$MHz bandwidth at various SNR levels with different estimation overheads.}
\end{figure}
In Fig.~7(a), the normalized spectral efficiency performance is evaluated for $125$MHz bandwidth with $512$ subcarriers using both auxiliary beam pair and grid-of-beams based angle estimation methods. Denote by $\bm{H}_{\mathrm{TR}}[k]=\bm{W}_{\mathrm{RF}}^{*}\bm{H}[k]\bm{F}_{\mathrm{RF}}$, we first define the conventional spectral efficiency metric as
\begin{equation}\label{sem}
R_{\mathrm{conv}}=\frac{1}{N}\sum_{k=0}^{N-1}\log_{2}\det\left(\bm{I}_{N_{\mathrm{S}}}+\frac{\gamma}{N_{\mathrm{S}}}\bm{H}_{\mathrm{TR}}[k]\bm{H}^{*}_{\mathrm{TR}}[k]\right).
\end{equation}
Specifically, $\bm{F}_{\mathrm{RF}}$ and $\bm{W}_{\mathrm{RF}}$ are constructed using the exact azimuth/elevation AoD and AoA for the perfect case, and estimated ones for the proposed algorithm and grid-of-beams based method. The angle mismatch and power imbalance are set to $\zeta=20^{\circ}$ and $\chi=0.2$. Further, $M_{\mathrm{tot}}=8$ and $N_{\mathrm{tot}}=64$ with $N_{\mathrm{x}}=4$ and $N_{\mathrm{y}}=8$ antenna elements on the $\mathrm{x}$ and $\mathrm{y}$ axes for each polarization domain are assumed. Denote the total number of coherence time slots for both the channel estimation and data communications by $T_{\mathrm{tot}}$, and the numbers of time slots used for auxiliary beam pair and grid-of-beams based angle estimation methods by $T^{\mathrm{ABP}}_{\mathrm{est}}$ and $T^{\mathrm{GoB}}_{\mathrm{est}}$. The normalized spectral efficiency is then defined as $(1-T^{\mathrm{ABP}}_{\mathrm{est}}/T_{\mathrm{tot}})R_{\mathrm{conv}}$ and $(1-T^{\mathrm{GoB}}_{\mathrm{est}}/T_{\mathrm{tot}})R_{\mathrm{conv}}$ for the proposed approach and grid-of-beams based design. We now interpret $T^{\mathrm{ABP}}_{\mathrm{est}}$ and $T^{\mathrm{GoB}}_{\mathrm{est}}$ using the number of iterations between the BS and UE for the proposed method and grid-of-beams based approach. As has been reported in \cite{singh2}, the computational complexity for the grid-of-beams based method is $E_{\mathrm{GoB}}=\left(N_{\mathrm{BM}}\right)^{N_{\mathrm{RF}}}\left(M_{\mathrm{BM}}\right)^{M_{\mathrm{RF}}}$, where $N_{\mathrm{BM}}$ and $M_{\mathrm{BM}}$ are the total numbers of candidate transmit and receive beams in the beam codebooks. Recall that for the proposed approach, the computational complexity is $E_{\mathrm{ABP}}=N_{\mathrm{RF}}N_{\mathrm{TX}}M_{\mathrm{RF}}M_{\mathrm{RX}}$ for the multi-RF case. Denote the maximum number of iterations between the BS and UE that can be supported in each time slot by $\epsilon_{\mathrm{t}}$. We then have $T_{\mathrm{est}}^{\mathrm{GoB}}=\lceil E_{\mathrm{GoB}}/\epsilon_{\mathrm{t}}\rceil$ and $T_{\mathrm{est}}^{\mathrm{ABP}}=\lceil E_{\mathrm{ABP}}/\epsilon_{\mathrm{t}}\rceil$. Here, we set $\epsilon_{t}=1000$ and $T_{\mathrm{tot}}=200$. Further, for simplicity, only the computational complexity in the azimuth domain search is accounted for within relatively small angular ranges, i.e., $[-15^\circ,15^\circ]$ for the BS and $[-45^\circ,45^\circ]$ for the UE. This is a reasonable assumption for a relatively large array size as the narrow beam search is generally executed at the final layer in the hierarchical beam search design \cite{wang,hur}. Here, we set $N_{\mathrm{BM}}=10$ and $M_{\mathrm{BM}}=4$ for the array size assumed above to cover the given angular ranges. Regarding $N_{\mathrm{S}}=N_{\mathrm{RF}}=M_{\mathrm{RF}}=\left\{2,3\right\}$, we set $N_{\mathrm{TX}}=\left\{20,30\right\}$ and $M_{\mathrm{RX}}=\left\{20,25\right\}$ for the proposed method. With different numbers of data streams $N_{\mathrm{S}}$, it can be observed from Fig.~7(a) that the auxiliary beam pair based method exhibits similar performance relative to that with perfect channel directional information at various SNR levels. Further, the proposed angle estimation algorithm shows superior performance relative to the grid-of-beams based approach in terms of the normalized spectral efficiency performance especially with relatively large $N_{\mathrm{S}}$. Similar observations can be obtained from Fig.~7(b), in which the normalized spectral efficiency performance is evaluated for $250$MHz bandwidth with $1024$ subcarriers and $N_{\mathrm{S}}=3$ under various values of $T_{\mathrm{tot}}$. It can be observed from Fig.~7(b) that for a relatively small $T_{\mathrm{tot}}$, the performance gap between the grid-of-beams based approach and the proposed method is significant as a large portion of $T_{\mathrm{tot}}$ is occupied for angle estimation in the grid-of-beams based approach.

\begin{figure}
\centering
\subfigure[]{%
\includegraphics[width=2.71in]{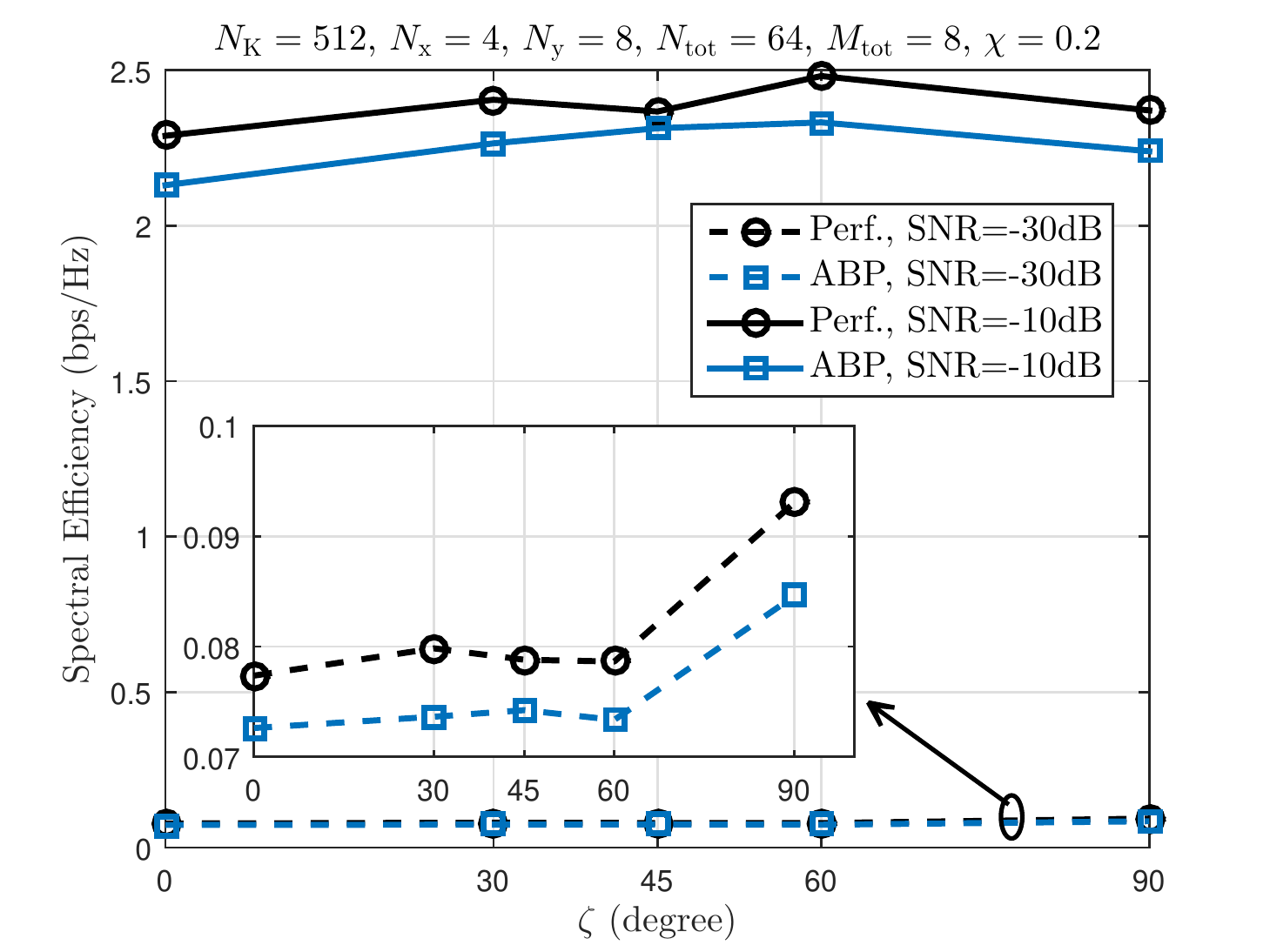}
\label{fig:subfigure3}}
\quad
\subfigure[]{%
\includegraphics[width=2.71in]{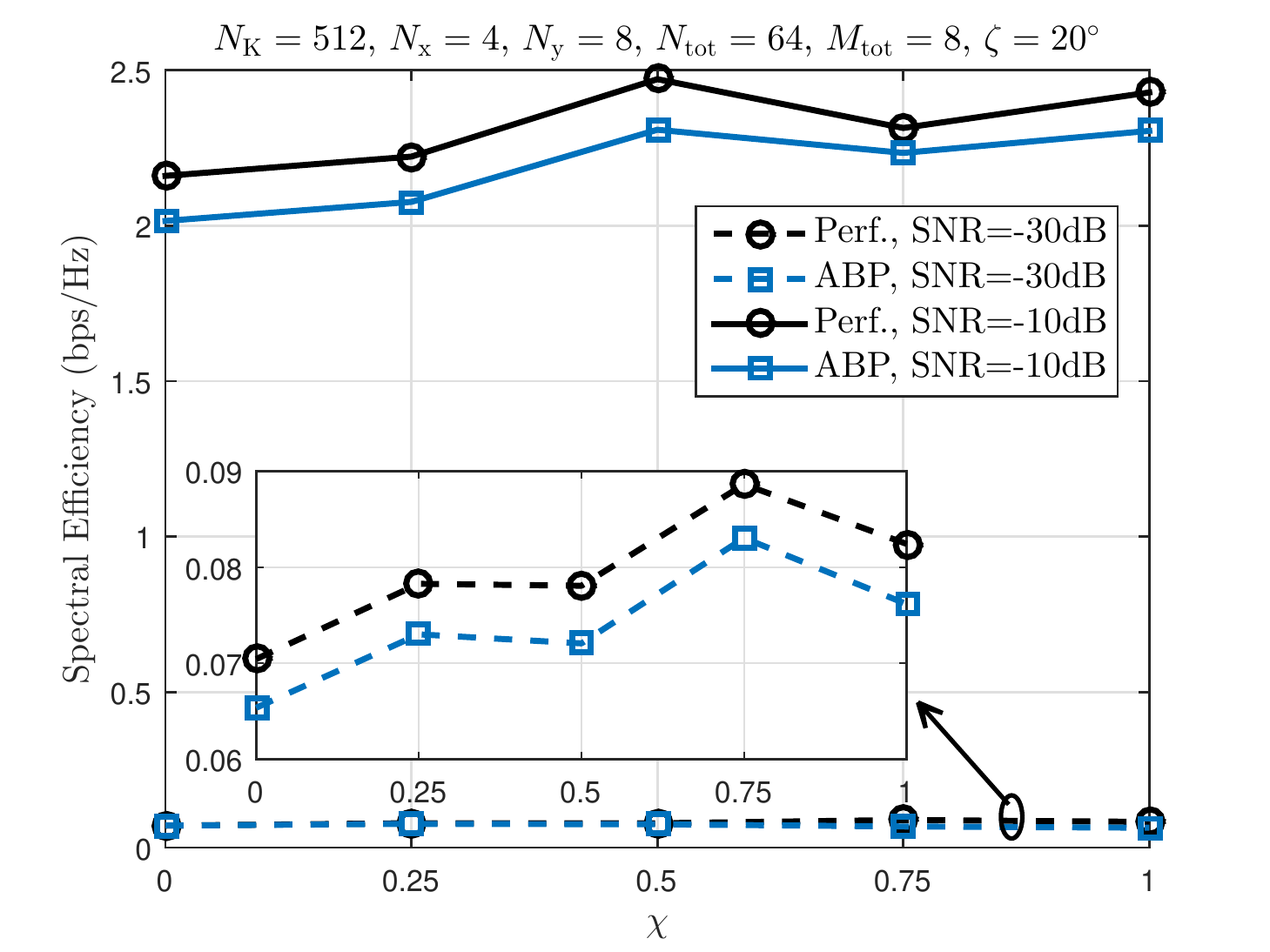}
\label{fig:subfigure3}}
\caption{(a) Spectral efficiency using perfect and estimated azimuth/elevation AoD and AoA via the proposed auxiliary beam pair design for $125$MHz bandwidth regarding various mismatched angles. (b) Spectral efficiency using perfect and estimated azimuth/elevation AoD and AoA via the proposed auxiliary beam pair design for $125$MHz bandwidth regarding various power imbalance values.}
\label{fig:figure}
\end{figure}
In Figs.~8(a) and 8(b), the effects of mismatched angle and power imbalance on the proposed design approach are evaluated in terms of the spectral efficiency calculated using (\ref{sem}). The main focus of the simulation plots provided in Figs.~8(a) and 8(b) is to verify the robustness of the proposed algorithm with respect to various mismatched angle and power imbalance assumptions. It can be observed from Fig.~8(a) that for different mismatched angles between the BS and UE antenna arrays, the performance gap between applying the proposed approach and that with perfect channel directional information is marginal. These observations are consistent with the asymptotical results derived in (\ref{asymttt}), (\ref{asymttx}) and (\ref{clra}). In Fig.~8(b), the spectral efficiency is plotted versus various power imbalance values. Similar to Fig.~8(a), the proposed auxiliary beam pair design is robust to variations in the power imbalance.
\section{Conclusion}
In this paper, we developed and evaluated an auxiliary beam pair based two-dimensional AoD and AoA estimation algorithm for wideband mmWave channels with cross-polarization. In the proposed design approach, by leveraging the well structured pairs of transmit/receive analog beams, high-resolution estimates of channel directional information can be obtained. We exposed several tradeoffs in our design including the mapping between the auxiliary beam pair and polarization, pilot design and feedback strategy. To evaluate our approach, we presented numerical results in a more elaborate channel than considered for the algorithm development. We found that good azimuth/elevation AoD and AoA estimation performance and effective rate could be achieved at various SNR levels, channel conditions and antenna configurations.

\bibliographystyle{IEEEbib}
\bibliography{main_bib_WCL}
\end{document}